\documentclass[12pt,thmsa]{article}
\usepackage{sw20lart}

\newcommand{\operatorname}[1]{\,\textrm{#1}\,}

\begin{document}

\author{Emilio Santos \and Departamento de F\'{i}sica. Universidad de Cantabria.
Santander. Spain \and email: santose@unican.es}
\title{Towards a picture of the natural world derived from relativity and quantum
theory}
\date{May 2026 }
\maketitle

\begin{abstract}
I contend that physics should provide a coherent account of reality, in
addition to be an efficient algorithm for the prediction of empirical
results. This article offers pictures of reality derived from theories of
modern physics. In particular it is shown that Bose quantum fields may be
interpreted as pure wave fields via the Weyl-Wigner representation. The most
relevant result being the existence of a stochastic vacuum field,
corresponding to the quantum vacuum fluctuations of the standard, canonical,
formulation of field theory. That field provides explanations for the
particle (photons) behaviour of the electromagnetic field. Also a realistic
interpretation is offered for the interference experiments with actual
particles, like atoms. The interpretation of classical general relativity is
standard but emphasis is made on the principle of equivalence. Problems like
the spacetime singularities (black holes) and the empirical violation of
Bell inequalities are touched but slightly. Asides from these problems the
main incompleteness of the article is the absence of a realistic
interpretation of Fermi fields.
\end{abstract}

\tableofcontents

\section{Introduction}

\subsection{The picture of reality}

The aim of this article is to offer a consistent view of physical reality
derived from relativity and quantum theory. I shall offer a personal opinion
that may disagree with common wisdom in several respects. Also the paper has
a limited scope, because dark points still remain. Nevertheless I hope that
the article may stimulate further advances in our view of nature.

The starting point is an epistemology of physics that I support, which was
well summarized in the initial paragraph of the celebrated EPR article:
``Any serious consideration of a physical theory must take into account the
distinction between the objective reality, which is independent of any
theory, and the physical concepts with which the theory operates. These
concepts are intended to correspond with the objective reality, and by means
of these concepts we picture this reality to ourselves''\emph{\ }\cite{EPR}.

The distinction between reality and concepts means that different
theoretical formalisms may exist for the description of a given domain of
reality. Therefore a specific theory, or a particular formalism, may be very
efficient in order to derive predictions for the results of experiments, but
other theories, or other formulations of the same theory, may be more
suitable to get a picture of reality. This is specially true with respect to
quantum theory, as I shall discuss in sections 3 to 7. Anyway I claim that
getting a picture of reality is an essential aspect of physics, or at least
a basic ingredient for a philosophical approach to the natural world.

\subsection{The difficulties for a realistic interpretation of quantum theory
}

For the sake of clarity about the purpose of this article, I shall start
with comments on the change in the views of the physical world with the
advent of relativity and quantum theories. Before the beginning of the 20th
century people believed in absolute space as the framework for all physical
processes, and also in absolute time that flows irreversibly, that is with a
future dramatically different from the past. In space there is matter which
consists of atoms, that is small particles which may remain bound, giving
rise to molecules, liquids or solids. The atoms have a random motion that we
call heat and it may explain statistically the laws of thermodynamics. The
motion of bodies under forces is governed by Newton\'{}s laws of dynamics.
Two fundamental forces were known: gravitational and electromagnetic,
formulated by Newton and Maxwell, respectively. The existence of
electromagnetic radiation was known, in particular it explained the nature
of light.

This world view, that defines ``classical physics'', is incomplete and in
several respects wrong. However it provided a picture of the physical world
that was \emph{clear}, that is free from internal contradictions. The
classical view changed dramatically in the first few years of the 20th
century. The theory of relativity modified our view of space and time.
Firstly they were unified in a spacetime, that was later declared to have
intrinsic curvature, and finally the curvature was tied to energy. The
relativistic world view, although strange or contrary to cherised
prejudices, still provided a clear picture of physical reality. Indeed
Einstein\'{}s relativity theory is considered a part of classical physics.

In sharp contrast, quantum theory appeared soon with dark aspects. In
particular from the beginning it involved a contradiction between the well
established wave theory of light and the corpuscle theory proposed by
Einstein in 1905. In fact this assumption, rather than Planck\'{}s quantum
of action, was the starting point for the problems of interpretation of
quantum theory. These problems did not diminish, rather they increased,
during the early years of quantum theory and culminated when Heisenberg
proposed his ``quantum mechanics'' ignoring any attempt to provide a picture
of reality, even rejecting it as misleading. In the rest of this subsection
I will further comment on the difficulties for the interpretation of quantum
theory.

Quantum mechanics is routinely used in laboratories with great success, but
no consensus on its interpretation has emerged \cite{Laloe}. Indeed many
interpretations have been proposed till now, as may be seen in a recent
Oxford Handbook about quantum interpretations \cite{Handbook}, which
contains articles of more than fifty authors supporting several different
views. In my opinion the absence of a consensus is a most important problem
in fundamental physics. Also I believe that having an interpretation that
enjoyed broad acceptance would be of great practical relevance, because its
absence slows down progress in several areas of research, e.g. quantum
information.

\emph{In this paper I seek for an interpretation of modern physics that
continues the tradition of classical physics. }Rather than attempting a
definition of that interpretation I shall clarify the subject with a
paradigmatic example: Newton\'{}s theory of the solar system. The theory
began with an intuitive model where the Sun is at the center and there are
rotating planets moving around it, that is the old \emph{heliocentric
hypothesis,} which became hegemonic after the observations and arguments of
Copernicus and Galileo. This simple model offers a qualitative understanding
of many observed facts: The turning of day and night, or summer and winter,
the eclipses, and sea tides. The model was complemented with the ascription
of numerical quantities to the various elements of the model, after the
careful observations of J. Kepler. The heliocentric model was a description,
or picture, of reality which became a physical theory when Newton achieved a
wonderful synthesis of the observed facts using his celebrated laws of
dynamics and gravity, respectively. They allow an accurate quantitative 
\emph{account} of the observations and measurements, with the possibility
of \emph{predicting} future events, like the dates of eclipses. In summary
the solar system theory consists of: 1) a picture of reality, and 2) Newton
physical laws.

The solar system theory may be compared with the quantum-mechanical atom,
where the theory provides just rules for the prediction of empirical results
but not a clear picture of reality. In fact our knowledge about the atom is
plagued with dark points. For the hydrogen atom the quantum formalism of
Schr\"{o}dinger provides a wavefunction, $\psi \left( \mathbf{r},t\right) ,$
but the actual meaning of $\psi $ is unknown. Max Born proposed that $\left|
\psi \right| ^{2}$ gives the probabilty \emph{of finding the electron at }$%
\left( \mathbf{r},t\right) $\emph{\ in a position measurement}. But
Born\'{}s proposal is just a rule that allows predictions about the result
of eventual human observations. Quantum mechanics is treated as an algoritm
for the prediction of empirical results, but it does not provide a \emph{%
clear picture of reality}. For instance it does not inform whether the
electron \emph{is }in a position at every time, or it \emph{becomes}
localized by the act of measuring the position.

Einstein was aware of the problem, as shown by a dialogue with Heisenberg
that took place in 1926 \cite{Heisenberg}. Einstein opened the conversation
with a question: ``What you have told us sounds extremely strange. You
assume the existence of electrons inside the atom...But you refuse to
consider their orbits''. The conversation continued for a while and, after
Heisenberg\'{}s arguments, Einstein warned: ``You are moving on very thin
ice. For you are suddenly speaking of what we know about nature and no
longer about what nature really does. In science we ought to be concerned
solely with what nature does''. Heisenberg arguments were the seed of the
Copenhagen interpretation of quantum mechanics, mainly supported by Niels
Bohr. It considers quantum mechanics as just an\ (extremely efficient in
fact) \emph{algorithm for the prediction of the results} of observations
or measurements, but it does not provides a picture of the world. In this
article I support the criticism of Einstein, and seek for a world picture
resembling the classical one, but resting on modern physics. In order to
give a short name to the seeked interpretion I will use the word
``realistic'', although I am aware that the choice of name may be
controversial. Anyway the word realism has become popular in the context of
the Bell inequalities, which may justify the election.

In my view there are two main difficulties for a satisfactory (realistic)
interpretation of quantum theory: 1) The abstract character of the standard
(``canonical'') formulation in terms of vectors and operators in Hilbert
space, and 2) The wave-particle duality. The solution of the former problem
is the use of an alternative formalim, which I propose to be the Wigner (or
Weyl-Wigner) representation for Bose quantum fields, see section 3.2 below.
However I have not yet a similar proposal for Fermi fields. Indeed I believe
that relativistic quantum fields are the fundamental inhabitants of our
world, while non-relativistic quantum mechanics is an approximations to
field theory. Therefore the interpretations of elementary quantum mechanics
should become later than that of fields, which will be the subject of
section 3.6 below. The wave-particle duality will be discussed in more
detail in section 6.

In summary, the mainstream of the physicists community believe that quantum
mechanics does not admit an interpretation resembling that of classical
physics. Nevertheless I am convinced that it is in fact possible. This
article discusses different cases where a realistic (classical-like)
interpretation is feasible. Unfortunately my proposal is incomplete, the
lack of a realistic interpretation for Fermi quantum fields being the most
relevant deficiency.

\subsection{Plan of the article}

This paper includes a review of my work on quantum interpretation from about
the year 2020 \cite{book}. I shall start with the interpretation of
(classical) relativity because it provides the theory of space and time,
which are the framework for all phenomena in the material world. Quantum
theory is currently perceived as more fundamental, whence people have
attempted treating (general) relativity in the framework of quantum theory,
for instance attempting to ``quantize gravity'', rather than trying to
develop a quantum field theory in the framework of curved spacetime. For
example assuming that spacetime should be treated as stochastic, the
randomness giving rise to a stochastic behaviour of the fields via the
Einstein equation (indeed my belief is that quantum fields are actually
stochastic fields). There are historical reasons for the perception of a
quantum supremacy. Firstly the great predictive power of quantum theory,
which has been tested more times and with higher precision than general
relativity. In particular quantum electrodynamics is the most accurately
tested theory of physics, leading to spectacular quantitative agreement with
empirical results. Secondly quantum theory is the basis for recent
technological developments, while the applications of general relativity are
almost restricted to astrophysics and cosmology. Anyway I think that quantum
theory must be studied in the framework of general relativity and not the
opposite way.

In the following section 2 I will offer the picture of space and time that,
in my view, emerges from relativity, and I shall postpone the interpretation
of quantum theory for later sections. Section 3.1 provides arguments
supporting my opinion that it is convenient to start with quantum fields,
rather than non-relativistic quantum mechanics, in order to obtain a picture
of reality. Thus, after a brief review of the Wigner representation in
section 3.2, and a digression on the canonical quantization, section 3.4
deals with a realistic interpretation of the quantized electromagnetic
field. In section 3.5 I shall discuss several consequences. In section 3.6 I
revisit the quantization of other Bose quantum fields, but it is pointed out
that a similar (realistic) approach does not yet exist for Fermi fields.
After that, the quantization of general relativity is discussed in section
4, and quantum particles in non-relativistic motion are studied in section
5. Section 6 deals with the realistic interpretation of the wave-particle
duality, in particular the phenomena that apparently prove the existence of
particles of light (``photons''). Finally in section 7 I discuss briefly the
phenomenon of quantum ``entanglement'' and comment on the Bell inequalities.
General conclusions of the article are presented in section 8.

Asides from the lack of interpretation for Fermi quantum fields and the
absence of a clear interpretation of quantum gravity, there are two
criticisms to this article which I foresee. Firstly in relation to the
matching of general relativity with quantum theory, the arguments offered in
the present paper would fail in some cases, namely when there are spacetime
singularities (in black holes). Secondly my view about non-local phenomena
(like the alleged empirical violation of the Bell inequalities) is contrary
to common opinion in the physicits community. These criticisms are sound and
I will comment on them in sections 4.3 and 7.2 respectively.

\section{Space and time in relativity theory}

\subsection{From Newton absolute space to the space-time of special
relativity}

The view of space and time has been dramatically modified with the advent of
relativity theory. Before the beginning of the XX Century the prevailing
opinion was that space and time were absolute realities, independent of
matter and of the means of observation. Contrary to that belief relativity
theory began with the assertion that both space and time should be treated
as means to describe properties of matter that depend on, or are relative
to, the observer. Hence the name ``relativity'' for the theory. It is
paradoxical that, in the final form of the theory i.e. general relativity,
space-time became the most fundamental reality of the material world,
although its geometry is closely tied to matter, this in the form of quantum
fields.

The question about absolute space became relevant when Newton formulated the
laws of mechanics, proposing that the acceleration is proportional to the
applied force. This led to the question: Acceleration with respect to what?.
Newton attempted to give an answer with the celebrated experiment of a
bucket filled with water. When the water rotates it is observed that the
center of the bucket is depleted with respect of the periphery. It was
observed that this centrifugal effect depends on the rotation of water, but
it is independent of the rotation, or not, of the bucket. This fact led
Newton to assume the existence of an ``absolute space'' with respect to
which the motion should be defined. (For a historical account with quotes by
Newton himself and other authors, including Einstein, see \cite{Newtonspace}%
). However the need of an absolute space may be avoided if we assume that
``the relevant acceleration is relative'' to the average mass of distant
matter (e.g. galaxies), as proposed by Ernst Mach \cite{Mach}. ``Mach
principle'' had a important influence on Einstein thinking \cite{Pais}.

Absolute time had been supported even more strongly, as a result of the
dramatic experience of human beings with the irreversible course of our
(human) life, from youth to old age. In Einstein\'{}s special relativity (of
1905) the absolute time was the most dramatic rejection. Indeed the theory
includes the possibility that the time lapse between two events, say A and
B, depends on the observer. Furthermore it may be that event A precedes
event B for some observer, but B precedes A for another observer in relative
motion with the former. However in 1908 Minkowski showed that the
mathematical structure of special relativity may be seen as the substitution
of a unified spacetime for the separated space and time. Events, i.e. points
in space-time, may be represented by four real numbers $\left\{
x,y,z,t\right\} $ with the property that for two different events, A and B,
both their distance $d_{AB}$ and time lapse $t_{AB},$ that is (assuming
Cartesian coordinates) 
\begin{equation}
d_{AB}=\sqrt{\left( x_{A}-x_{B}\right) ^{2}+\left( y_{A}-y_{B}\right)
^{2}+\left( z_{A}-z_{B}\right) ^{2}},t_{AB}=t_{B}-t_{A},  \label{d}
\end{equation}
may be ``relative'', but there is a quantity, the interval, which is
``absolute'', independent of the observer . It is defined by

\begin{equation}
I=\sqrt{\left| d_{AB}^{2}-t_{AB}^{2}c^{2}\right| },  \label{I}
\end{equation}
where $c$ is the speed of light. Space and time were unified as spacetime,
mathematically a pseudo-Euclidean variety with four dimensions. In the words
of Einstein ``physics became the study of a 4-dimensional object, rather
than the evolution of a 3-dimensional object'' \cite{Pais}. The spacetime
defined so far (that is, according to \emph{special} relativity) is named
``Minkowski space''. Relevant properties of that space are summarized as
follows.

The free (inertial) motion corresponds to the position, $\bf r$, of a
particle changing in proportion to time t, that is 
\begin{equation}
\mathbf{r=r}_{0}\mathbf{+v}t,\left| \mathbf{v}\right| \leq c,
\label{inertial}
\end{equation}
where c is the speed of light in vacuum. The existence of an upper limit to
the possible velocities is one of the main consequences of relativity
theory. Eq.$\left( \ref{I}\right) $ implies that for a body in inertial
motion, in particular at rest, the following quantity defined between two
given events (points in Minkowski space) A and B, is invariant 
\begin{equation}
\tau _{AB}\equiv \sqrt{t_{AB}^{2}-d_{AB}^{2}c^{-2}}=t_{AB}\sqrt{1-\frac{v^{2}%
}{c^{2}}},  \label{propertime}
\end{equation}
that is the same for all inertial observers. It is named ``proper time'' and
it corresponds to the time measured by an observer at rest with respect to
the body. The time $t_{AB}$ seen by another inertial observer is never
smaller than the proper time $\tau _{AB}$.

Proper time may be defined more generally, for any body with arbitrary
motion between events A and B placed at a distance $d_{AB},$ via the
integral 
\begin{equation}
\tau _{AB}=\int_{t_{A}}^{t_{B}}\sqrt{1-\frac{\left| \mathbf{v}(t)\right| ^{2}%
}{c^{2}}}dt,d_{AB}=\left| \int_{t_{A}}^{t_{B}}\mathbf{v}(t)dt\right| .
\label{propert}
\end{equation}
A consequence of this definition is that the proper time $\tau _{AB}$
between two given spacetime points is a maximum for an observer moving with
constant velocity, $\mathbf{v}$, that is when the motion is inertial.

For a correct understanding of general relativity a historical digression is
convenient, which is made in the following.

\subsection{In search for a relativistic field theory of gravity}

The origin of general relativity was the search for a relativistic theory of
gravity. In fact Newton\'{}s law 
\begin{equation}
F_{grav}=G\frac{Mm}{r^{2}},  \label{G}
\end{equation}
does not fit in special relativity, indeed it predicts instantaneous actions
at a distance. Then searching for a relativistic theory of gravity was one
of the main scientific goals after Einstein\'{}s special relativity of 1905.
In order to see the difficulties for that aim let us consider the
formulation of Newton\'{}s gravity as a field theory.

The concept of \emph{field} \emph{of force} had been introduced by M.
Faraday around 1840 in electromagnetism in order to avoid the problem of
actions at a distance that had troubled Newton. Applying the concept of
field to gravitation, from eq.$\left( \ref{G}\right) $ we may get the
gravitational potential $\phi \left( \mathbf{r}\right) $ and the field
intensity $\mathbf{g}\left( \mathbf{r}\right) $ created by a mass
distribution $\rho \left( \mathbf{r}\right) ,$ as follows 
\begin{equation}
\phi \left( \mathbf{r}\right) =-G\int d^{3}\mathbf{r}^{\prime }\rho \left( 
\mathbf{r}^{\prime }\right) \left| \mathbf{r}^{\prime }-\mathbf{r}\right|
^{-1},\mathbf{g}\left( \mathbf{r}\right) =-\bigtriangledown \phi \left( 
\mathbf{r}\right) ,  \label{g1}
\end{equation}
where $\bigtriangledown \equiv \left( \partial /\partial x,\partial
/\partial y,\partial /\partial z\right) .$ The gravitational force $\mathbf{f%
}$ on a (pointlike) test particle with mass $m$ placed at $\mathbf{r}$ would
be 
\begin{equation}
\mathbf{f}_{grav}\mathbf{=}m\mathbf{g}\left( \mathbf{r}\right) .  \label{fg}
\end{equation}
The acceleration of the particle by the gravity force changes its kinetic
energy, whence conservation of total energy implies that the field itself
should possess energy, which may be transfered to the particle. In fact we
must ascribe a \emph{negative} energy density $\rho _{grav}\left( \mathbf{r%
}\right) $ to the gravitational field as follows 
\begin{equation}
\rho _{grav}\left( \mathbf{r}\right) =-\frac{1}{2G}\left| \mathbf{g}\left( 
\mathbf{r}\right) \right| ^{2}.  \label{g2}
\end{equation}

We might compare the passage, from Newton\'{}s non-relativistic theory of
gravity to a relativistic theory, with the route from the non-relativistic
Coulomb law of electrostatic to the relativistic Maxwell theory of
electromagnetism. Of course that process required a lot of experimental and
theoretical work that lapsed for most the 19th century. Also Maxwell theory
preceded Einstein special relativity, but conceptually the latter may be
considered the framework of the former.

The relevant point is that Newton theory of gravity as well as both Coulomb
and Maxwell theories are linear in the sense that the effects (attraction or
repulsion forces) are proportional to the causes (masses or electric charges
and currents, respectively). However any relativistic generalization of
Newton gravitational theory could not be linear. In fact special relativity
establish that energy $E$ is related to mass $m$ via 
\begin{equation}
E=mc^{2},  \label{mc}
\end{equation}
$c$ being the speed of light. Therefore a relativistic theory of gravity
should take into account the (negative) mass-energy of the field eq.$\left( 
\ref{g2}\right) .$ Hence the gravitational field $\mathbf{g}\left( \mathbf{r}%
\right) ,$ eq.$\left( \ref{g1}\right) ,$ should be replaced by the following
one 
\begin{equation}
\mathbf{g}\left( \mathbf{r}\right) =-\int d^{3}\mathbf{r}^{\prime }\left| 
\mathbf{r}^{\prime }-\mathbf{r}\right| ^{-1}\left[ G\rho \left( \mathbf{r}%
^{\prime }\right) -c^{-2}\left| \mathbf{g}\left( \mathbf{r}^{\prime }\right)
\right| ^{2}\right] ,  \label{g3}
\end{equation}
where the latter term comes from the energy of the field itself. Actually
that term is usually negligible whence eq.$\left( \ref{g1}\right) $ may be a
good approximation for eq.$\left( \ref{g3}\right) .$ However from a
fundamental point of view that term also gravitates, whence the equation
should be modified again giving rise to the appaerance of another term of
order $c^{-4}$, this would give rise to a term of order $c^{-6},$ and so on.
This shows the non-linear character of the wanted relativistic gravity
theory, which gives rise to big difficulties.

We may compare eq.$\left( \ref{fg}\right) $ with the electrostatic force
from another point of view. Writing the electric force in the form 
\begin{equation}
\mathbf{f}_{electric}\mathbf{=}q\mathbf{E}\left( \mathbf{r}\right) ,
\label{Coulomb}
\end{equation}
where $q$ is the electric charge of a particle and $\mathbf{E}\left( \mathbf{%
r}\right) $ the electric field. In both cases, electrostatic and
gravitational, the equation of motion for a particle should be Newton second
law of mechanics, that is 
\begin{equation}
\mathbf{f}=m\mathbf{a,}  \label{E}
\end{equation}
where $\mathbf{a}$ is the acceleration. The point is that in the electric
case the force depends on the electric charge $q,$ but the acceleration in
terms of the force involves a \emph{different} parameter, the mass $m$. In
sharp contrast, in the gravitational case the mass appears in both the force
eq.$\left( \ref{fg}\right) $ and the law of motion eq.$\left( \ref{E}\right)
,$which looks something strange. This peculiar fact may be stated saying
that all bodies experience the same acceleration in a gravitational field,
because from eqs.$\left( \ref{fg}\right) $ and $\left( \ref{E}\right) $ we
may get 
\begin{equation}
\mathbf{a=-g.}  \label{equivalence}
\end{equation}
In particular all bodies fall with the same acceleration near the Earth
surface (neglecting perturbations by the air), as had been discovered by
Galileo and was essential for Newton proposal of a \emph{universal}
gravitational law eq.$\left( \ref{G}\right) .$ It is universal because it
governs both the motion of celestial bodies, like that of moon around earth,
and the motion of bodies near the earth surface.

In 1907 Einstein realized that eq.$\left( \ref{equivalence}\right) $ might
be a fundamental law of nature, that is ``the gravity field is equivalent to
an acceleration'', which is known as \emph{Principle of equivalence.}
Einstein interpreted the equivalence principle as stating that: 1) the
motion under gravity is just the ``natural free motion'', i.e. purely
inertial, which explains Galileo\'{}s discovery, and 2) the inertia at a
point is determined by the distribution of masses around it. Einstein
considered the equivalence principle, and his intepretation, to be the
``gl\"{u}cklichste Gedanke meines Lebens'' (the happiest though of my life) 
\cite{Pais}. The principle requires, or strongly suggests, that spacetime is
curved, whence inertial motion ceases to be a straight line with constant
velocity, which might explain the motion of bodies under ``gravity''. In the
years from 1907 to 1915 Einstein studied Riemann theory of manifolds with
curvature, worked hard and finally arrived at general relativity, ``the most
beautiful theory of physics'' in the words of Lev Landau.

\subsection{Spacetime and matter in general relativity}

General relativity (GR) introduced 3 innovations: 1) It reinforced the
relevance of spacetime as a fundamental framework, 2) It led to the need of
spacetime curvature, and 3) It strongly tied spacetime with matter. These
innovations gave rise to the picture that follows.

At a difference with the (flat) Minkowski spacetime of special relativity,
in GR the spacetime is a variety with intrinsic curvature \cite{Weinberg},
as described in the following. Events, i.e. points in space-time, may be
represented by four coordinates, the real numbers $\left\{
x^{1},x^{2},x^{3},x^{4}\right\} .$ Then curvature may be derived from the
expresion of the infinitesimal interval $ds$ amongst two close events in
terms of the coordinates, that is

\begin{equation}
ds^{2}=\sum_{\mu =1}^{4}\sum_{\nu =1}^{4}g_{\mu \nu }\left(
x^{1},x^{2},x^{3},x^{3}\right) dx^{\mu }dx^{\nu },  \label{metric}
\end{equation}
which is known as \emph{the metric}. Actually $ds$ corresponds to an
infinitesimal proper time as defined in eq.$\left( \ref{propertime}\right) .$

The method to determining curvature via the metric was introduced by the
mathematician Carl F. Gauss for surfaces. The points of a surface may be
determined by 2 coordinates (\emph{x,y), }and the metric $dl,$ which gives
the infinit\'{e}simal distance between two points, may be written in terms
of these coordinates, that is 
\[
dl^{2}=A\left( x,y\right) dx^{2}+B\left( x,y\right) dy^{2}+C\left(
x,y\right) dxdy, 
\]
to be compared with eq.$(\ref{metric})$ for 4 coordinates of spacetime. From
the functions \emph{A, B, C}, Gauss derived a single number, named
curvature of the surface at the point (x,y), which corresponds to $1/R^{2},R$
being the radius of the sphere most close to the surface near the said
point. For instance if \emph{A} and \emph{B} are constant (independent
of x,y) and C=0, Gauss curvature is nil (i.e. $R\rightarrow \infty )$ and
the surface is said flat. Bernhard Riemann generalized Gauss theory for
varieties with N dimensions, and Einstein used Riemann theory for the
4-dimensional spacetime.

In Riemann\'{}s theory, combining the elements of the metric tensor $g_{\mu
\nu },$ eq.$\left( \ref{metric}\right) ,$ with their first and second
derivatives respect to the coordinates, it is possible to get the Riemann
tensor $R_{\mu \nu \lambda \sigma }$, whose nil value is a necessary and
sufficient condition for zero curvature, the spacetime being flat
(Minkowski) in that case. On the other hand matter (here the word matter
includes radiation) is characterized by quantities like energy, momentum and
angular momentum, that may be summed up via an energy-momentum tensor $%
T_{\mu \nu }^{matt}$. The fundamental equation of general relativity,
Einstein\'{}s equation, relates curvature with the energy-momentum tensor of
matter and radiation $T_{\mu \nu }^{matt}$, that is 
\begin{equation}
R_{\mu \nu }-\frac{1}{2}g_{\mu \nu }R\equiv G_{\mu \nu }=-8\pi GT_{\mu \nu
}^{matt},  \label{Einst}
\end{equation}
where $G$ is Newton constant of gravity. The left side $G_{\mu \nu },$ named
Einstein tensor, involves the Ricci tensor $R_{\mu \nu }$ and the scalar $R$%
, that may be derived from the Riemann tensor, that is 
\begin{equation}
R_{\mu \nu }=\sum_{\lambda \sigma }g^{\lambda \sigma }R_{\lambda \mu \sigma
\nu },R=\sum_{\mu \nu }g^{\mu \nu }R_{\mu \nu }.  \label{Ricci}
\end{equation}

A popular understanding of eq.$\left( \ref{Einst}\right) $ may be summarized
with \emph{Wheeler slogan}: ``Spacetime tells matter how to move, matter
tells spacetime how to curve''. Indeed spacetime curvature is characterized
by Einstein tensor $G_{\mu \nu }$ and matter (including radiation) by the
energy-momentum tensor $T_{\mu \nu }^{matt}.$ In the next section I shall
propose a conceptual simplification of this view of general relativity.

I stress again that \emph{gravity is not a force} according to GR, but the
acceleration of the bodies motion is a consequence of spacetime curvature.
In fact, the inertial motion corresponds to a path in spacetime which may be
parametriced by a parameter $s$, via four functions $x_{j}\left( s\right) ,$
j=1,2,3,4 with the condition that the \emph{proper time} is a maximum
between an initial event, $s=s_{a},$ and a final one, $s=s_{b},$that is 
\[
s_{ab}\equiv \int_{s_{a}}^{s_{b}}ds=\textrm{ maximum.} 
\]
This is a generalization to curved space of the condition stated after eq.$%
\left( \ref{propert}\right) $ in flat (Minkowski) space. The motion that
maximizes proper time may be labeled a \emph{geodesic }in spacetime\emph{%
. }Thus free motion follows spacetime geodesics\emph{.}

Let us comment on the fact that Einstein tensor is nil in regions without
matter, but Riemann tensor may be finite (not nil) in those regions. In fact
the nil value of the Riemann tensor $R_{\lambda \mu \sigma \nu }$ is a
necessary and sufficient condition for the absence of curvature, but in
contrast the nil value of Einstein tensor $G_{\mu \nu }$ is necessary but
not sufficient. The reason for that is a peculiar property of the Riemann
curvature in 4 dimensions, namely the Riemann tensor may be written as a sum
of two tensors as follows (see e.g. \cite{Weinberg} ) 
\begin{eqnarray}
R_{\lambda \mu \sigma \nu } &=&M_{\lambda \mu \sigma \nu }+C_{\lambda \mu
\sigma \nu },  \nonumber \\
M_{\lambda \mu \sigma \nu } &=&\frac{1}{2}\left( g_{_{\lambda \sigma
}}R_{\mu \nu }-g_{_{\lambda \nu }}R_{\mu \sigma }-g_{_{\mu \sigma
}}R_{\lambda \nu }+g_{\mu \nu }R_{\lambda \sigma }\right)  \nonumber \\
&&-\frac{R}{6}\left( g_{_{\lambda \sigma }}g_{_{\mu \nu }}-g_{_{\lambda \nu
}}g_{_{\mu \sigma }}\right) .  \label{Weyltensor}
\end{eqnarray}
$C_{\lambda \mu \sigma \nu }$ is named Weyl tensor and it does not
contribute to the Einstein tensor$,$ that is $G_{\mu \nu }$ may be got
putting $M_{\lambda \mu \sigma \nu }$ in place of $R_{\lambda \mu \sigma \nu
}$ in eq.$\left( \ref{Ricci}\right) ,$ as may be easily checked. This fact
may be stated saying that the curvature (Riemann) tensor is in part
energy-momentum (that is a linear function of the Ricci tensor) and in part
``gravitation'' in itself (Weyl tensor). This may be seen putting ec.$\left( 
\ref{Einst}\right) $ in eq.$\left( \ref{Weyltensor}\right) ,$ which gives

\begin{eqnarray}
M_{\lambda \mu \sigma \nu } &=&-4\pi G\left( g_{_{\lambda \sigma }}T_{\mu
\nu }-g_{_{\lambda \nu }}T_{\mu \sigma }-g_{_{\mu \sigma }}T_{\lambda \nu
}+g_{\mu \nu }T_{\lambda \sigma }\right)  \nonumber \\
&&+\frac{4GT}{3}\left( g_{_{\lambda \sigma }}g_{_{\mu \nu }}-g_{_{\lambda
\nu }}g_{_{\mu \sigma }}\right) ,  \label{M}
\end{eqnarray}
where G is Newton constant, $T_{\mu \nu }$ the momentum-energy tensor, and $%
T=\sum g^{\rho \tau }T_{\rho \tau }.$

This allows a nice interpretation for the strange ``action at a distance''
of gravity. The pre-relativistic solution to the problem was the
introduction of the concept of field of force eq.$\left( \ref{g2}\right) $,
but it gave rise to difficulties as commented in section 2.2. In sharp
contrast in GR the Weyl tensor appears naturally when Einstein eq.$\left( 
\ref{Einst}\right) $ is solved (integrated). Therefore the curvature, and
consequently the motion of matter, is determined by the Weyl tensor in
regions without matter (nor radiation) where the tensor $T_{\mu \nu }$ is
nil and consequently also $M_{\lambda \mu \sigma \nu }=0$. This is the case
for instance in the exterior of spherical bodies like the Earth where the
inertial motion consists of elipses that are geodesics in spacetime, but the
Einstein tensor $G_{\mu \nu }$ is nil there. We may say that the Weyl tensor
plays in GR the role of the gravitational field in Newtonian theory, but in
GR it appears naturally while in Newtonian gravity it is an \emph{ad hoc }%
supplement.

Integrating eq.$\left( \ref{Einst}\right) $ means getting the metric
elements as functions of the coordinates, that is determining the functions $%
g_{\mu \nu }\left( x_{1},x_{2},x_{3},x_{4}\right) $ for some region defined
by appropriate boundary conditions (e.g. the whole space assuming that
curvature goes to zero at infinity). The integration requires that the
metric tensor $g_{\mu \nu }$ should be twice derivable respect to the
coordinates $\left\{ x^{1},x^{2},x^{3},x^{4}\right\} .$ (In view of this
requirement for the derivability of the metric tensor it is not strange that
Einstein was reluctant to believe in the existence of actual singularities 
\cite{Pais}, which nevertheless at present are assumed to exist in black
holes, see section 3.6.3 below).

\subsection{Is general relativity a field theory of gravity?}

In Maxwell electromagnetic theory the field appears as something real. The
typical example is light, which is just an electromagnetic field. It travels
from the source to the detector (say from the Sun to the eyes of people on
Earth). From Maxwell time the relevance of fields has increased and the
current belief is that the inhabitants of the universe are Relativistic
Quantum Fields. Therefore Quantum Field Theory (QFT) has become the
fundamental theory of nature. It has led to predictions which have achieved
a truly spectacular agreement with empirical data, specially in domain of
Quantum Electrodynamics. This fact has given rise to the widespread opinion
that general relativity should be quantized like all other known fields of
the standard theory of fundamental particles, whence the attempt of treating
GR as the relativistic field theory of gravity. I do not agree, I believe
that GR is not a field theory of gravity, but the denial that gravity is a
field. It is a theory of spacetime and its relation with matter.

In my view GR might be a field theory of gravity if the quantities $g_{\mu
\nu }$ were interpreted as potentials of the field, this property being
unrelated in principle to spacetime curvature. If this were the case it
would be plausible to have a different tensor determining the metric. This
possibility has been studied for instance by A. A. Logunov in his
``Relativistic theory of gravity'' \cite{LogunovM}. I believe that the
interpretion of Einstein eq.$\left( \ref{Einst}\right) $ with the double
role of determining the spacetime curvature and being a gravitational field
is unnecesary and destroys the conceptual simplicity of GR. My point of view
is that GR is quite different from the relativistic fields of high energy
physics.

It is true that the study of systems with strong energy density requires a
joint treatment of spacetime curvature and quantum features, as is the case
in spacetime singularities (in black holes and the very early universe).
This has given rise to a research programme known as ``quantum gravity''.
However I believe that what is needed is an appropriate quantization of
spacetime, as I will discuss in section 4, but not a quantization of
gravity, because ``gravity'' is just the name given to the fact that
spacetime has intrinsic curvature, a fact that modifies the evolution of
fields.

\subsection{A geometrical view of mechanical quantities}

General relativity allows connecting the field equations with the spacetime
curvature via three steps. Firstly there are equations that provide the
mass-energy contents of the fields in the form of an energy-momentum tensor $%
T_{\mu \nu }.$ Then eq.$\left( \ref{Einst}\right) $ relates that tensor with
Einstein\'{}s $G_{\mu \nu }$. Finally the mathematical theory of Riemann
relates Einstein tensor with the metric tensor. I argue that we can make a
conceptual simplification reducing the steps from 3 to 2. We cannot remove
the former and the latter steps, but we may remove the second step because
Einstein equation is just an equality of two tensors modulo the Newton
constant $G$. Thus it is enough to take Einstein equation as an \emph{%
identity}, rather than an \emph{equality} relating different concepts in
order to reduce the mentioned steps from 3 to 2. With that view eq.$\left( 
\ref{Einst}\right) $ might be seen as a kind of ``dictionary'' that
translates from mechanical language (energy-momentum tensor) to geometrical
language (Einstein tensor). The dimensional (Newton) constant is needed
because, for historical reasons, we use units of length (and time when the
speed of light is put c=1) in the metric tensor, therefore in Einstein
tensor too, which are different from the units used in the energy-momentum
tensor (mass, energy, momentum or pressure).

Of course I do not mean that Einstein equation is a trivial discovery. It
involves the highly non-trivial assumption that spacetime is curved, which
strongly changes our view of the world. What I mean is that GR allows a
beautiful \emph{geometrical interpretation} of the mecanical concepts like
energy, momentum and angular momentum. Along the historical development of
physics people introduced these concepts which were fundamental for the
development of physics, but now we have a charming interpretation for them
thanks to Einstein equation. I argue that they are forms of spacetime
curvature.

The proposal to change the interpretation of eq.$\left( \ref{Einst}\right) $
from equality to identity may seem a mere semantic issue, but it allows a
better understanding of our world. Firstly it makes more compelling the
opinion that there are not 4 fundamental forces in nature, just 2 because
gravity is not a force and electromagnetism is unified with the weak
interactions. Or only 1 if electroweak force were unified with the strong
interaction. Secondly it simplifies the picture of the world because it
reduces the number of concepts needed for the description, the dynamical
variables not being primitive but geometrical concepts.

The picture that emerges is that the world consists of fields in a curved
spaccetime. Possibly also particles, but I shall exclude particles at this
stage, see below. The numerical values of the fields (possibly with several
components each) at every point in a region are constrained by spacetime
curvature, the field equations and boundary conditions. On turn the fields
determine the Einstein tensor via the field equations in a region of
spacetime. Every Einstein tensor is associated to a class of curvature, the
class consisting of all Riemann tensors giving the same Ricci tensor or,
equivalently, the same energy-momentum tensor.

This picture corresponds to what is named ``classical general relativity''.
In order to discuss the ``quantization of general relativity'' (QGR) it is
necesary to study previously the interpretation of quantum theory, which
will be the subject of section 3. Quantization of GR will be discussed in
section 4, but previously I shall coment on the irreversibility of time, in
the subsection 2.6 which follows.

\subsection{Irreversibility in the universe}

Although marginal for this paper I will comment briefly on the so-called
arrow of time. From ancient times people believed that space does not
possess any special direction, it is the presence of Earth what causes the
difference between vertical and horizontal directions. More common was the
belief on an essential direction of time distinguishing past from future.
Physicists however were reluctant to admit that this ``arrow of time'' is
fundamental. As Einstein stated: ``For us convinced physicist the
distinction between past and future is an illusion, although a persistent
one'' \cite{EinstBesso}. This is supported by the fact that no violation has
been found of the product of the 3 fundamental discrete symmetries, that is
CPT, that could provide a truly fundamental physical arrow of time. Of
course it is true that the fulfillement of CPT in the fundamental laws of
physics is compatible with the violation of C, P and T symmetries.

Thus the observed arrow of time required an explanation. Boltzmann
statistical proof that a closed system evolves spontaneously toward
equilibrium was an important achievement which elucidated irreversibility in
many cases, but it cannot provide an explanation as to why ``we grow old'',
something recognized at Boltzmann\'{}s time. The true reason for the
existence of a general arrow of time on Earth, and hence the irreversibility
of our life, derives from the fact that the Earth receives energy from the
Sun at high temperature (about 6000 K) and reemits it at low temperature
(about 300 K) thus producing an increase of entropy. On turn this is a
consequence of the expansion of the universe, that leads to irreversible
star evolution. Thus the appropriate explanation of the irreversibility on
Earth has become after the discovery that the universe is expanding.

\section{Quantum theory. The Weyl-Wigner representation}

\subsection{Lack of consensus on the interpretation. The realistic approach}

At the end of the 19th century there was an apparently well established
picture of matter, resting on the physics known at that time. This state of
affairs was dramatically altered by quantum theory. In fact quantization was
not conceived as the substitution of a new picture of reality for the
classical image, but the withdrawal of the old picture without the advent of
a new one. Indeed quantum mechanics was proposed by Heisenberg with an
explicit resignation to pictures of reality.

The problem remains until today. In fact no consensus on its interpretation
has emerged \cite{Handbook}. The result is that, after one century of
quantum mechanics, we find ourselves in a strange situation. Everybody who
has learned quantum mechanics agrees how to use it but we do not understand
the meaning of this strange conceptual apparatus that each of us uses so
effectively to deal with our world.

The initial formulation of the theory was a nonrelativistic quantum
mechanics of particles (QM in the rest of this section). Popular approaches
to interpret quantum theory usually begin with QM, with the hope of
extending later the interpretation to relativistic quantum theory. I think
that this is an error because the quantum particles studied in QM cannot be
treated as similar to classical particles. In fact, from a fundamental point
of view it is not obvious why the nonrelativistic approximation of
interacting relativistic quantum fields may be treated as a set of particles
under the action of the electromagnetic field, as usual. For instance
electrons and nuclei in the quantum theory of atoms, molecules and solids.
However the particles are under the action of fields (including their
vacua), these fields being hidden in the nonrelativistic approximation. In
my opinion the action of the fields is essential for the quantum behaviour
of the particles. Indeed in section 5 below I will present a model where the
introduction of a non-local potential may explain the wave behaviour of
particles (e.g. interference of electrons, neutrons or atoms). In my view
that potential simulates the action of the hidden fields, in the
non-relativistic approximation. In summary the quantum particles in the
non-relativistc approximation are actually rather complex objects consisting
of interacting fields dresssing bare particles. Thus I believe that in order
to find a realistic interpretation of quantum theory it is suitable to try
understanding firstly the (relativistic quantum)\ fields, rather than the
quantum mechanics of non-relativistic particles.

The purpose of Heisenberg was to formulate QM with ingredients as close as
possible to measurable quantities. Thus his quantum mechanics substituted
arrays of numbers, e.g. frequencies and intensities of atomic spectra, for
the dynamical variables of classical mechanics like position, momentum or
energy. The said arrays of numbers (matrices in mathematical language) may
be multiplied with each other, but the product is not commutative. Dirac
replaced the matrices by abstract vectors and operators in a linear space
which, after mathematical elaboration by J. von Neumann, is defined as a
Hilbert space. The result is an elegant formalism which has become the 
\emph{canonical formulation of quantum theory}. However it does not offer
an intuitive picture of reality, at a difference with classical physics.
Thus QM looks like an (efficient) algoritm for the prediction of empirical
results, whose physical interpretation is dark.

The alternative Schr\"{o}dinger\'{}s wave mechanics appeared, after the work
of L. de Broglie, as an attempt to unify two images popular in classical
physics, but incompatible with each other: particles (localized) and waves
(extended). Schr\"{o}dinger\'{}s initial proposal of understanding his
``wavefunction'' as a continuous distribution of mass or electric charge
failed, because the localized (particle) behaviour of electrons was proved
necessary in order to understand many-electron atoms. Max Born introduced a
practical interpretation of the wavefunction assuming that its modulus
square is a probability density for the position of a particle. From that
time on the \emph{probability amplitude} (e.g. for the positions of
particles) has been the cornerstone of the whole quantum theory. However in
my view ``probability amplitude'' is a strange union of unrelated words
which many people admits as an explanation, e.g. of wave-particle duality,
but I do not.

In summary none of the two initial formulations of QM leads naturally to a
picture of physical reality. The consequence has been the early supremacy of
Heisenberg-Bohr (Copenhagen) pragmatic approach, that values the predictive
power of the theory but rejects physical pictures as misleading. Later on
many other interpretations have been proposed \cite{Handbook}, but none
fully satisfactory for most people.

I am convinced that: 1) A (realistic) interpretation providing a picture of
reality is possible, 2) It cannot be achieved from the canonical (Hilbert
space) formalism, but from the Wigner representation, maybe with some
modifications, 3) We should start with the interpretation of quantum fields,
4) After that we might get a picture of the mechanics of quantum particles.

As a consequence, in the following I shall firstly revisit the formalism
initiated by Weyl and Wigner which leads to a realistic interpretation of
quantum Bose fields, in particular electromagnetism (in sections 3.2 to
3.5). There is not yet a similar formulation for Fermi fields, whence our
interpretation will be incomplete. For non-relativistic quantum motion a
realistic interpretation is provided in section 5 below, but it is
convenient to deal previously with spacetime quantization which will be made
in section 4.

\subsection{ Weyl-Wigner formalism for quantum particles}

In 1932 Wigner introduced a new formalism for quantum mechanics which has a
classical flavour \cite{Wigner}. He proposed the following representation, $%
W_{\psi }(x,p)$, for the state of a particle with wavefunction $\psi (x)$
(in one dimension for simplicity): 
\begin{equation}
W_{\psi }(x,p)\equiv \frac{1}{\pi 
\rlap{\protect\rule[1.1ex]{.325em}{.1ex}}h%
}\int \psi (x+y)\psi (x-y)\exp \left( 2ipy/
\rlap{\protect\rule[1.1ex]{.325em}{.1ex}}h%
\right) dy,  \label{Wigner}
\end{equation}
which is named Wigner function of the quantum state. The generalization to N
particles with 3N coordinates and 3N momenta is straightforward. It is
possible also to define a Wigner representation for observables, whence both
states and observables become functions in phase space, that is the space of
the coordinates $\left\{ x_{j}\right\} $ and momenta $\left\{ p_{j}\right\} $
of the particles. The expectation values $\left\langle M\right\rangle _{f}$
are obtained via integrals like 
\begin{equation}
\left\langle M\right\rangle _{f}=\int M\left( \left\{ x_{j},p_{j}\right\}
\right) f\left( \left\{ x_{j},p_{j}\right\} \right) \Pi _{j}dx_{j}dp_{j},
\label{9}
\end{equation}
where $f$ and $M$ represent the state and the observable, respectively. The
numerical values of these expectations agree with those got from the
previous formulations, in particular the canonical, Hilbert space, formalism 
\cite{Scully}, \cite{Zachos}. As a result the Wigner representation is a
different formalism for the same physical theory, that is quantum mechanics.

The proof of equivalence may be most easily seen via the Weyl transform,
which leads to the Wigner representation starting from the canonical
(Hilbert space) formalism, rather than from (Schr\"{o}dinger\'{}s) wave
mechanics as in eq.$\left( \ref{Wigner}\right) $. Weyl introduced his
transform in 1927 as a method of quantization \cite{Weyl}. We may suppose
the following naive argument at the origin of Weyl transform. For a relation
in classical mechanics, e.g. 
\[
F\left( x,p\right) =0, 
\]
we might find a similar relation involving operators, $F^{\prime }(\hat{x},%
\hat{p}),$ via the transform 
\begin{equation}
F^{\prime }\left( \hat{x},\hat{p}\right) =\int F\left( x,p\right) \delta
\left( x-\hat{x}\right) \delta \left( p-\hat{p}\right) dxdp,  \label{10}
\end{equation}
where $\delta \left( {}\right) $ are Dirac deltas. (From now on I will label
operators by a ``hat'', e.g. $\hat{x},\hat{p}).$ Eq.$\left( \ref{10}\right) $
is a symbolic expression, not a sensible mathematical equation, because
Dirac delta is defined for numerical arguments, but not for operators. In
order to give a meaning to eq.$\left( \ref{10}\right) $ we may substitute
integral representations for the deltas. That is (modulo appropriate
regularization) 
\begin{equation}
\delta (x-y)=\frac{1}{2\pi }\int_{-\infty }^{\infty }\exp \left[ i\lambda
\left( x-y\right) \right] d\lambda .  \label{11}
\end{equation}
This equality is valid for numerical $x$ and $y$, but the integrand on right
side is meaningful even if $x$ and/or $y$ are operators. Hence we might
substitute integral representations for the deltas in eq.$\left( \ref{10}%
\right) $. However a difficulty remains because the operators $\hat{x}$ and $%
\hat{p}$ do not commute whence the integral representation associated to the
symbolic expression $\delta \left( x-\hat{x}\right) \delta \left( p-\hat{p}%
\right) $ is different from that associated to $\delta \left( p-\hat{p}%
\right) \delta \left( x-\hat{x}\right) .$

Weyl proposed a transform which leads from classical functions like $F(x,p)$
to (quantum) functions $F_{sym}\left( \hat{x},\hat{p}\right) .$ The subindex 
$sym$ means symmetrical order, that is writing any product involving
operators in all possible orderings and dividing by the number of terms, for
instance 
\begin{equation}
(\hat{x}^{2}\hat{p})_{sym}=\frac{1}{3}\left( \hat{x}^{2}\hat{p}+\hat{x}\hat{p%
}\hat{x}+\hat{p}\hat{x}^{2}\right) .  \label{12}
\end{equation}
Weyl transform may be written (for a single particle in one dimension) 
\begin{equation}
F_{W}\left( \hat{x},\hat{p}\right) =\frac{1}{4\pi ^{2}}\int F(x,p)\exp
\left[ i\lambda \left( x-\hat{x}\right) +i\mu \left( p-\hat{p}\right)
\right] d\lambda d\mu ,  \label{13}
\end{equation}
whose generalization to many particles in 3D is straightforward.

Most interesting in the following is the inverse Weyl transform, which may
be written as follows 
\begin{equation}
F(x,p)=\frac{1}{4\pi ^{2}}\int d\lambda d\mu Tr\left\{ \hat{F}^{\prime }\exp
\left[ i\lambda \left( \hat{x}-x\right) +i\mu \left( \hat{p}-p\right)
\right] \right\} ,  \label{14}
\end{equation}
where and $\hat{F}^{\prime }$ is an operator and $Tr$ means the Trace
operation. It is not difficult to prove that if we write $\hat{F}^{\prime }$
in the form $\left| \psi \rangle \langle \psi \right| $ eq.$\left( \ref{14}%
\right) $ leads to eq.$\left( \ref{Wigner}\right) ,$ which proves the
equivalence with the Wigner representation (that in the following I will
name Weyl-Wigner, WW for short). The inverse Weyl transform allows getting,
from states and observables in the canonical (Hilbert space) formalism, the
corresponding states and observables in WW. The expectation values
calculated in WW, eq.$\left( \ref{9}\right) ,$ agree with those calculated
in the canonical formalism, whence WW represents the same physical theory
than the canonical formulation in terms of Hilbert spaces, because both
predict the same measurable quantities (expectation values).

The Wigner representation has a classical flavor, but the classical
appearance is misleading. In fact, the Wigner functions $f\left( \left\{
x_{j},p_{j}\right\} \right) $ are not positive definite in general, whence
the states cannot be interpreted as probability distributions in phase
space. Therefore the Wigner representation is currently seen as just a
useful calculational tool for some specific problems (of quantum statistical
mechanics in particular), but it does not offer a realistic interpretation
of quantum mechanics of particles.

In summary the formalism of nonrelativistic QM, plus the rules named
``measurement theory'', provide a good tool for the prediction of the
results of experiments, but it does not give clues for a picture of reality.
Therefore QM is not a good starting point in order to achieve a realistic
interpretation of quantum theory.

\subsection{Difficulties for a picture of reality of the canonical quantized
fields}

I shall deal with the electromagnetic field or, more generally, Bose fields.
The standard method to describe a classical field is to expand it in plane
waves or, more generally, normal modes. For instance in the simple casse of
a scalar (Bose) field the expansion in plane waves may read 
\begin{equation}
\phi (\mathbf{r},t)=\sum_{\mathbf{k}}\phi _{\mathbf{k}}\equiv \sum_{\mathbf{k%
}}\sqrt{\frac{2\pi 
\rlap{\protect\rule[1.1ex]{.325em}{.1ex}}h%
}{\omega V}}\left[ a_{\mathbf{k}}\exp \left( i\mathbf{k\cdot r-}i\omega
_{l}t\right) +a_{\mathbf{k}}^{*}\exp \left( -i\mathbf{k\cdot r+}i\omega
t\right) \right] .  \label{expansion}
\end{equation}
The amplitudes of the modes are conveniently written using two complex
conjugate quantities (c-numbers) $\left\{ a_{j},a_{j}^{*}\right\} ,$ where $%
j $ labels a mode. The standard (canonical) quantization method consists of
promoting the amplitudes to be operators $\left\{ \hat{a}_{j},\hat{a}%
_{j}^{\dagger }\right\} $ with appropriate commutation rules, that is 
\begin{equation}
\hat{a}_{j}\hat{a}_{k}^{\dagger }-\hat{a}_{k}^{\dagger }\hat{a}_{j}=\delta
_{jk},\hat{a}_{j}\hat{a}_{k}-\hat{a}_{k}\hat{a}_{j}=\hat{a}_{j}^{\dagger }%
\hat{a}_{k}^{\dagger }-\hat{a}_{k}^{\dagger }\hat{a}_{j}^{\dagger }=0,
\label{com}
\end{equation}
where $\delta _{jj}=1,\delta _{jk\neq j}=0$. The evolution of the field
operators $\left\{ \hat{a}_{j}\left( t\right) ,\hat{a}_{j}^{\dagger }\left(
t\right) \right\} $ has formal similarity with the motion of mechanical
harmonic oscillators, which may be shown performing the change of variables

\begin{equation}
\hat{x}_{j}\equiv \frac{c}{\sqrt{2}\omega _{j}}\left( \hat{a}_{j}+\hat{a}%
_{j}^{\dagger }\right) ,\hat{p}_{j}\equiv \frac{i 
\rlap{\protect\rule[1.1ex]{.325em}{.1ex}}h%
\omega _{j}}{\sqrt{2}c}\left( \hat{a}_{j}-\hat{a}_{j}^{\dagger }\right) ,
\label{aj}
\end{equation}
where $
\rlap{\protect\rule[1.1ex]{.325em}{.1ex}}h%
$ is Planck constant, $c$ the velocity of light and $\omega _{j}$ the
frequency of the normal mode.

The vacuum state is represented either by the state vector $\mid
vac\rangle $ or by the density operator 
\begin{equation}
\hat{\rho}=\mid vac\rangle \langle vac\mid ,  \label{2d}
\end{equation}
fulfilling 
\[
\hat{a}_{j}\mid vac\rangle =0=\langle vac\mid \hat{a}_{j}^{\dagger }, 
\]
for any annihilation operator $\hat{a}_{j},$ $0$ being here the nul vector
in the Hilbert space.

Excited pure states of the radiation field in the canonical formalism are
associated to vectors in the Hilbert space that are obtained by repeated
application of the creation oparators to the vacuum state. Thus a generic
pure state may be represented as follows 
\begin{equation}
\hat{f}^{\dagger }\mid 0\rangle ,  \label{stateHS}
\end{equation}
where $\hat{f}^{\dagger }$ means any polynomial of the creation operators $%
\left\{ \hat{a}_{j}^{\dagger }\right\} .$ Mixed states are probability
distributions of pures states, represented by density operators. The
observables, the expectation values and the evolution are well known and I
shall skip them.

Up to here the formalism, now I will pass to the interpretation. We might
try to interpret the formalism as representing a wave field with a somewhat
strange representation by vectors and operators in a Hilbert space. However
the standard opinion is not this, but the formalims is assumed to represent
both, waves and particles. Indeed relativistic quantum fields are supposed
to represents a kind of particle each, every particle corresponding to a
definite spin $s$ and mass $m$. For instance for the electromagnetic (EM)
field $s=1,m=0$. In fact the operators $\left\{ \hat{a}_{j}^{\dagger
}\right\} $ are believed to create particle states, photons in the case of
the EM field. For instance the vector (of the Hilbert space) $\hat{a}%
_{j}^{\dagger }\mid 0\rangle $ is a one-photon state. However this
assumption gives rise to a difficult problem of interpretation. In fact, the
state represented by $\hat{a}_{j}^{\dagger }\mid 0\rangle $ will be
associated to a plane wave, in the expansion of the field, corresponding to
one of the terms of eq.$\left( \ref{expansion}\right) $. But it may be a
spherical wave for another choice of expansion. Thus the picture of the
boson offered by $\hat{a}_{j}^{\dagger }\mid 0\rangle $ depends on our
choice of expansion. Furthermore, bosons (in particular photons) are
intuitively associated to (small) particles rather than to extended plane
(or spherical) waves.

In summary, any clear interpretation of a quantum Bose field as consisting
of waves and particles is not possible. Hence most people supports the
belief that quantum field theory cannot provide a picture of reality, and
even it should not. I strongly disagree. I am convinced that the picture is
possible, but as wave fields only. The particle behaviour would be an effect
of the vacuum field. In my view the picture is crystal clear using the WW
representation rather than the canonical, Hilbert space, formalism.

\subsection{Weyl-Wigner formulation of the quantum electromagnetic field}

\subsubsection{The formalism}

The WW formalism, developed for particles, may be extended to fields in
particular the electromagnetic (EM) field. In the following I provide a
short review of the formalism, details may be seen elsewhere \cite{Universe1}%
. The Weyl transform eq.$\left( \ref{14}\right) $ leads from the operators
eqs.$\left( \ref{aj}\right) $ to classical-like amplitudes. This fact allows
a straightforward formulation of the electromagnetic field, in the WW
representation \cite{Universe1} , which may be extended to other Bose fields 
\cite{Universe2}.

As in the WW formalism for particles, the Weyl transform eq.$\left( \ref{14}%
\right) $ allows deriving the product of amplitudes in WW for any product of
creation or annihilation operators in the canonical (Hilbert space, HS)
formalism. In fact if we have a symmetric product of operators like $\left( 
\hat{a}_{j}^{m}\hat{a}_{j}^{\dagger n}\right) _{sym}$ in the canonical (HS)
formalism, the WW counterpart is 
\begin{equation}
\left( \hat{a}_{j}^{m}\hat{a}_{j}^{\dagger n}\right) _{sym}\rightarrow
a_{j}^{m}a_{j}^{*n}  \label{sym}
\end{equation}
where \emph{sym} stands for symmetric and it means writing a sum of the $%
m+n$ operators in all possible orderings and then dividing by the number of
terms, that is $\left( m+n\right) !/(m!n!)$. If the product of operators in
the canonical (Hilbert space, HS) formalism is not symmetric it is possible
to get a symmetric expresion using the commutation rules eqs.$\left( \ref
{com}\right) $. Simple examples of the transform from HS to WW are 
\begin{equation}
\hat{a}_{j}^{\dagger }\hat{a}_{j}\rightarrow \left| a_{j}\right| ^{2}-\frac{1%
}{2},\hat{a}_{j}^{\dagger }\hat{a}_{j}\hat{a}_{j}^{\dagger }\hat{a}%
_{j}\rightarrow \left| a_{j}\right| ^{4}-\left| a_{j}\right| ^{2}.
\label{sym1}
\end{equation}

The vacuum state in WW may be got inserting the density operator $\hat{\rho}%
, $ eq.$\left( \ref{2d}\right) $, in place of $\hat{M}$ in eq.$\left( \ref
{14}\right) $. We get, after some algebra,

\begin{equation}
W_{0}=\prod_{j}\frac{2}{\pi }\exp \left( -2\left| a_{j}\right| ^{2}\right) ,
\label{1}
\end{equation}
which is normalized for the integration with respect to $\prod_{j}d\operatorname{Re}%
a_{j}d\operatorname{Im}a_{j}.$ Taking the Lorentz invariance of the vacuum field
into account, the mean energy $E_{j}$ associated to a field component of the
vacuum field in the expansion eq.$\left( \ref{expansion}\right) $ is
proportional to $\omega _{j}$ and it is neccessary to choose $\left\langle
E_{j}\right\rangle $ $=\frac{1}{2}
\rlap{\protect\rule[1.1ex]{.325em}{.1ex}}h%
\omega _{j}$ \cite{Milonni}. This fact was already taken into account for
the choice of coefficients in eq.$\left( \ref{expansion}\right) .$ Hence eq.$%
\left( \ref{1}\right) $ leads to 
\begin{equation}
W_{0}=\prod_{j}\frac{2}{
\rlap{\protect\rule[1.1ex]{.325em}{.1ex}}h%
\omega _{j}}\exp \left( -\frac{2E_{j}}{
\rlap{\protect\rule[1.1ex]{.325em}{.1ex}}h%
\omega _{j}}\right) ,  \label{W0}
\end{equation}
where $E_{j}$ is the mean energy of mode $j$, the normalization is
appropriate for integration with respect to $\prod_{j}dE_{j}$.

The WW\ counterparts of the states eq.$\left( \ref{stateHS}\right) $ may be
obtained taking the Weyl transform eq.$\left( \ref{14}\right) $ into
account, which gives for every one of these states a function of the
amplitudes $\left\{ a_{j}\right\} $ named Wigner function of the state. Thus
the Wigner functions of the canonical (HS) pure states are as follows 
\begin{equation}
W_{\psi }\left( \left\{ a_{j}\right\} \right) =T_{W}\left( \hat{f}^{\dagger
}\mid 0\rangle \langle 0\mid \hat{f}\right) ,  \label{stateWW}
\end{equation}
where $\hat{f}^{\dagger }\mid 0\rangle \langle 0\mid \hat{f}$ is the density
operator corresponding to the state vector eq.$\left( \ref{stateHS}\right) $
and $T_{W}$ means the (inverse) Weyl transform eq.$\left( \ref{14}\right) .$

\textbf{Observables} are represented by functions of the electric and
magnetic field, which are related to the observables in the canonical (HS)
formalism via the expansion in normal modes followed by the Weyl transform
eq.$\left( \ref{14}\right) .$ The expectation value of an observable in a
state is obtained via the integral of the product of the corresponding
functions of the amplitudes.

Eq.$\left( \ref{14}\right) $ allows getting the WW counterparts of the
observables in the HS formalism. In particular the free field Hamiltonians
are, respectively, 
\begin{equation}
\hat{H}_{HS}=
\rlap{\protect\rule[1.1ex]{.325em}{.1ex}}h%
\sum_{j}\omega _{j}(\hat{a}_{j}^{\dagger }\hat{a}_{j}+\frac{1}{2})=\frac{1}{2%
}
\rlap{\protect\rule[1.1ex]{.325em}{.1ex}}h%
\sum_{j}\omega _{j}(\hat{a}_{j}^{\dagger }\hat{a}_{j}+\hat{a}_{j}\hat{a}%
_{j}^{\dagger }),H_{WW}=
\rlap{\protect\rule[1.1ex]{.325em}{.1ex}}h%
\sum_{j}\omega _{j}\left| a_{j}\right| ^{2}.  \label{HamiltoniansW}
\end{equation}
However in the canonical formalism it is common to change the order of the
operators putting the annihilation to the right, which is known as `normal
ordering rule'. Using that rule the canonical and WW Hamiltonians become,
respectivly, 
\begin{equation}
\hat{H}_{HS}^{normal}=
\rlap{\protect\rule[1.1ex]{.325em}{.1ex}}h%
\sum_{j}\omega _{j}\hat{a}_{j}^{\dagger }\hat{a}_{j},H_{WW}^{normal}= 
\rlap{\protect\rule[1.1ex]{.325em}{.1ex}}h%
\sum_{j}\omega _{j}\left( \left| a_{j}\right| ^{2}-\frac{1}{2}\right) ,
\label{HnormalW}
\end{equation}
where I have taken eq.$\left( \ref{sym1}\right) $ into account$.$ Hence the
vacuum energy is \emph{defined} to be zero in HS, but this choice requires
a reinterpretation in WW as shown below, see comments on eq.$\left( \ref
{photocounts}\right) .$

\textbf{Expectation values} in the canonical formalism read $Tr(\hat{\rho}%
\hat{M})$, or in particular $\langle \psi \mid \hat{M}\mid \psi \rangle ,$
and the translation to the WW formalism leads to the integral of the product
of two functions of the amplitudes, that is 
\begin{equation}
Tr(\hat{\rho}\hat{M})=\int W_{\hat{\rho}}\left\{ a_{j},a_{j}^{*}\right\} W_{%
\hat{M}}\left\{ a_{j},a_{j}^{*}\right\} \prod_{j}d\mathrm{Re}a_{j}d\mathrm{Im%
}a_{j},  \label{expect}
\end{equation}
where $W_{\hat{\rho}}$ and $W_{\hat{M}}$ are the counterparts of a density
operator, $\hat{\rho},$ and a quantum observable $\hat{M}.$ A particular
case of eq.$\left( \ref{expect}\right) $ is the vacuum expectation value
where $W_{\hat{\rho}}$ becomes $W_{0}.$

The \textbf{evolution }of the states in the WW formalism is given by the
Moyal equation, 
\begin{eqnarray}
\frac{\partial W}{\partial t} &=&\frac{2}{
\rlap{\protect\rule[1.1ex]{.325em}{.1ex}}h%
}\sum_{n=0}^{3N}\frac{\left( -1\right) ^{n}}{\left( 2n+1\right) !}\left[ 
\frac{
\rlap{\protect\rule[1.1ex]{.325em}{.1ex}}h%
}{2}\left( \frac{\partial }{\partial x_{j}}\frac{\partial }{\partial
p_{j}^{\prime }}-\frac{\partial }{\partial p_{j}}\frac{\partial }{\partial
x_{j}^{\prime }}\right) \right] ^{2n+1}  \nonumber \\
&&\times \left[ W\left( \left\{ x_{j}\mathbf{,}p_{j}\right\} \right)
H_{part}\left( \left\{ x_{j}^{\prime }\mathbf{,}p_{j}^{\prime }\right\}
\right) \right] \equiv \left\{ W,H_{part}\right\} _{M},  \label{dW}
\end{eqnarray}
where we should identify $\left\{ x_{j}^{\prime },p_{j}^{\prime }\right\}
=\left\{ x_{j},p_{j}\right\} $ after performing the derivatives \cite{Zachos}
, \cite{book}. $\left\{ W,H_{part}\right\} _{M}$ is named Moyal bracket. For
simplicity I have written it in terms of canonical variables of mechanics,
but the evolution eq.$\left( \ref{dW}\right) $ is valid also for the
radiation field if we perform a c-number change of variables similar to eq.$%
\left( \ref{aj}\right) .$

In the case of the free EM field, the Hamiltonian $H_{WW}$ eq.$\left( \ref
{HamiltoniansW}\right) $ is quadratic in the amplitudes whence terms with $%
n\neq 0$ do not contribute to eq.$\left( \ref{dW}\right) .$ Only terms
without Planck constant remain. Then Moyal bracket becomes Poisson\'{}s ,
which proves that \emph{the evolution of the quantized free EM field in
the WW formalism is just the classical (Maxwell) evolution }\cite{EPJP}.

Furthermore the interaction Hamiltonian $H_{int}$ of the field with a system
of charged particles is given in terms of the potential vector, which is
also linear in the field amplitudes $\left\{ a_{j},a_{j}^{*}\right\} $.
However $H_{int}$ is \emph{not }quadratic in the coordinates and momenta
of the particles in general, whence the evolution of the particles is not
classical. In particular the Moyal eq.$\left( \ref{dW}\right) $ for the
particles depends on Planck constant $
\rlap{\protect\rule[1.1ex]{.325em}{.1ex}}h%
$, which however does not appear in the field evolution. For details see 
\cite{EPJP}.

\subsubsection{Realistic interpretation}

Up to here the mere WW formalism. In order to make a realistic
interpretation possible we must introduce several assumptions which do not
follow from the Weyl transform. The most relevant refers to the states. In
fact, the Weyl transform of the states, as defined in the standard
formalism, cannot be states in WW if we want a realistic interpretation. For
instance, as is well known the Weyl transform of a single-photon state (the
Wigner function of the state) is not positive definite, therefore it could
not be considered a physical state in a realistic interpretation of the WW
formalism.

In general eq.$\left( \ref{stateWW}\right) $ suggests an interpretation of
the Wigner function $W_{\psi }\left( \left\{ a_{j}\right\} \right) $ as a
probability distribution of amplitudes. However a necessary condition for
this interpretation would be that $W_{\psi }$ is nonnegative definite and
normalized. The latter constraint holds provided that the statevector eq.$%
\left( \ref{stateHS}\right) $ is normalized, which I assume. However the
positivity puts a problem because there are many canonical states eq.$\left( 
\ref{stateHS}\right) $ whose Wigner function is not positive. Thus the set
of states in WW does not fit in the set of states of the canonical
formalism. Indeed we shall assume that physical states of the field in WW
should correspond to radiation with ( positive) probability distribution of
field amplitudes $\left\{ a_{j},a_{j}^{*}\right\} $ \emph{superposed to
the vacuum field ZPF}. I believe that this difficulty does not prevent a
realistic interpretation of the WW formalism for the quantized EM field.

Eqs.$\left( \ref{1}\right) $ and $\left( \ref{W0}\right) $ strongly suggest
an interpretation of the vacuum state $W_{0},$ of the quantized EM field in
the WW formalism, as a probability distribution of amplitudes.\emph{\ Thus
it suggests a picture of the quantum vacuum as a real random radiation, a
stochastic field, filling space with the distribution eq.}$\left( \ref{W0}%
\right) .$ \emph{That random radiation has a mean energy }$\frac{1}{2} 
\rlap{\protect\rule[1.1ex]{.325em}{.1ex}}h%
\omega _{j}$\emph{\ per normal mode, and it is currently named zeropoint
field (ZPF).}

In spite of the very different defintion of states in either the canonical
formalism or WW, \emph{I conjecture that all experiments, in the domain of
validity, which may be interpreted with the canonical formalism might be
also interpreted within WW}. Indeed my belief about the general
interpretation of quantum theory may be put as follows: The ``measurement
theory'' is an addition which, although useful for some calculations, should
not be taken as an essential part of the theory. Here the measurement theory
includes the definitions of states and observables. In my view choosing the
appropriate ``quantum state'' representing a preparation and the adequate
``quantum observable'' representing an observation or measurement, is a
difficult task that should be carefully studied in every actual experiment.

I shall finish the section elucidating why the WW formalism is appropriate
for the EM field but not for non-relativistic quantum mechanics of particles
(QM)?. Indeed in both cases the set of states in WW does not fit in the set
of canonical (Hilbert space, HS) formalism, due to a requirement of
positivity in the former which is not demanded in HS. The response is that
the situation is quite different in both cases, for the following reasons:

1. In quantized EM field (QEM) the ground state has a Wigner function, eq.$%
\left( \ref{1}\right) ,$ which is positive definite. In QM the ground state
of a system of particles has frequently a Wigner function not positive.

2. The evolution of the free QEM field is governed by the classical
Maxwell-Lorentz laws, which preserve positivity of probability
distributions. This is not the case in QM, governed by Moyal eq.$\left( \ref
{dW}\right) .$

3. The states of QEM are most times produced by the action of macroscopic
devices on the vacuum, which would give rise to states with positive Wigner
function. This happens for instance in spontaneous parametric down
conversion leading to entangled photon pairs, see section 7.1 below.

4. In the canonical (HS) interpretation of experiments, n-photon states
usually appear but at intermediate stages of the calculation, which may not
be positive when translated to the WW formalism via eq.$\left( \ref{14}%
\right) .$ However there is no need to ascribe physical reality to these
intermediate (mathematical) elements of the calculation.

5. In typical QED calculations the ``photon propagator'' involves the 
\emph{vacuum} expectation value and that state has a positive Wigner
function.

\subsection{Weyl-Wigner formalism for quantum electrodynamics}

In subsections 3.4.1 and 3.4.2 I have reviewed the properties of the
electromagnetic field alone, but quantum electrodynamics would involve also
the charges. The combination of the quantized EM field with Fermi fields,
that is relativistic QED, cannot be treated within the WW formalism because
we do not have an appropriate (realistic) interpretation of Fermi fields.
The combination of the EM field with non-relativistic particles cannot be
treated within WW too, because the Wigner representation for those particles
does not admit a realistic interpretation in general as commented above.

\subsubsection{Quantum optics}

A domain where the field interacts with matter and both may be treated
within the WW formalism is quantum optics, when the EM field interacts with
macroscopic bodies. Indeed macroscopic bodies may be treated within
classical electrodynamics, and the combination with the quantized EM field
gives precisely a study within the WW formalism. In this case the WW
treatment corresponds to just classical Maxwell-Lorentz electrodynamics with
the addition a zeropoint Gaussian random field in the vacuum \cite{MS}, \cite
{book}. In some cases that treatment has given rise to things, like
``negative probability distributions'', that cannot admit a realistic
interpretation \cite{Scullyoptics}. However, I am convinced that all
problems might be eliminated with a careful treatment within WW. 

Actually the phenomena within quantum optics that present greatest
difficulties for a realistic interpretation are photon entanglement and the
optical tests of Bell inequalities, which are discussed in Sections 7.1 and
7.2. In the following I recall 3 simple but interesting examples which have
been studied in more detail elsewhere \cite{Foundations}. Further examples
will be provided in section 6 devoted to the wave-particle duality.

\subsubsection{Casimir effect}

It is the attraction between two parallel perfectly conduction plates placed
in vacuum. The reason of the force is that the plates restrict the possible
modes of the radiation field because in equilibrium the component of the
electric field parallel to the plate surface should be nil. Thus the vacuum
energy of the ZPF with the plates in place is different from the energy with
the plates removed, and the dependence of the energy with the distance
between plates gives rise to a force. The calculation in the WW formalism is
closely related to the standard one in HS \cite{Milonni}. The picture that
we get in WW is that the pressure of the ZPF is different on the two sides
of each plate, which gives rise to the force.

\subsubsection{Atoms in cavities}

As shown in the theory of the Casimir effect the ZPF radiation modes in
confined space are different from the modes in free space. Then, assuming
that spontaneous emission is partially stimulated by the ZPF, the lack of
some radiation modes would prevent that excited atoms decay emitting
radiation in the said modes. In fact the experiments have shown the
inhibition of atomic decays in cavities giving rise to increased lifetimes
of excited atoms. For a semiclassical model see \cite{Humberto}.

\subsubsection{Stability of matter. Ground state of the hydrogen atom\ }

The atom cannot be studied within the WW formalism because it involves
charged particles, not just the field. This means that the predictions
obtained may disagree with quantum mechanics. However it is the case that it
provides a realistic interpretation for some properties of the atom. The
interpretation is interesting because it suggests an explanation for the
stability of matter.

In a simplified model the hydrogen atom consists of two particles, proton
and electron, characterized each by the mass and the electric charge. The
proton mass being much larger than the electron mass we may study the atom
assuming that the proton is at rest. In classical \emph{mechanics} the
electron may move around the nucleus, say in a circle having energy $E$ and
we might write the following equalities 
\begin{equation}
\left| E\right| =\frac{1}{2}mv^{2}=\frac{1}{2}\frac{e^{2}}{r},v=r\omega ,
\label{03}
\end{equation}
According to classical \emph{electrodynamics} the electron would radiate
leading to a collapse of the atom but, taking the ZPF into account as a real
random radiation, the atom may also absorb energy from the field. The
combination of emission and (random) absorption perturbes the motion which
would be irregular, not circular. However it is plausible that eqs.$\left( 
\ref{03}\right) $ are roughly fulfilled on the average. A dynamical
equilibrium may arise when the atomic kinetic energy becomes equal to the
energy of the radiation mode having the same frequency, that is $\left|
E\right| \sim \frac{1}{2}
\rlap{\protect\rule[1.1ex]{.325em}{.1ex}}h%
\omega $ . Indeed this equality is plausible because the electron will
interact most strongly with such modes. Hence the energy and the size of the
atom may be got removing the quantities $v$ and $\omega $ from eqs.$\left( 
\ref{03}\right) $, which leads to 
\begin{equation}
E\sim -\frac{me^{4}}{2
\rlap{\protect\rule[1.1ex]{.325em}{.1ex}}h%
^{2}},r\sim \frac{
\rlap{\protect\rule[1.1ex]{.325em}{.1ex}}h%
^{2}}{me^{2}},  \label{04}
\end{equation}
in agreement with the quantum prediction and experiments.

\subsection{Realistic interpretation of Bose fields via the WW formalism}

I believe that Bose fields are continuous fields similar to the
electromagnetic one, the particle behaviour being an effective property, see
section 4 below. In contrast I have not a clear picture of Fermi fields, but
I conjecture that they might consist of a sea of particles and antiparticles
similar to the original picture of Dirac.

Thus we might assume that in nature there are bare particles with a definite
mass and charge, but the physical particles have quite different values for
those quantities, as shown by the need of renormalization techniques in
quantum electrodynamics. The reason for the difference is the ``dressing''
due to many quantum fields including their vacua. For instance a physical
electron should be seen as an extended object with size of order the Compton
wavelength, having observable mass and charge far from the bare quantities.
The electron position might be defined by the center of the charge
distribution, but the momentum and angular momentum would involve
substantial contributions from fields. Hence, at a difference with classical
mechanics, the electron state cannot be represented by a point in phase
space. In particular its future evolution is not determined by just initial
position and momentum. For this reason I do not propose to get a picture of
QM using the Weyl-Wigner formalism, which nevertheless does yield a fair
realistic interpretation for the EM field, as studied in section 4. For Bose
fields other than electromagnetism the WW treatment is similar. Thus WW
strongly suggests a realistic interpretation for the quantized Bose fields 
\cite{Universe2}.

In summary, there are several formalisms for the study of quantum systems
and in this article we deal with two of them: canonical (Hilbert space, HS)
and Weyl-Wigner (WW). Both are valid as calculational tools with different
efficiency, the canonical one being most useful in general. Only one, WW,
provides a physical realistic interpretation for Bose fields and none of
them for Fermi fields or nonrelativistic quantum mechanics, as was commented
on section 3.

In fact we do not have a transform for Fermi fields that would play the role
of Weyl transform for Bose fields. Thus our realistic interpretation of
quantum fields is not yet complete. In my view getting an appropriate
formalism for Fermi fields would be a dramatic improvement for the
interpretation of quantum theory. It would permit a realistic interpretation
of the whole quantum field theory. For the evolution of quantum particles in
non-relativistic motion there is a formalism allowing a realistic
interpretation, presented in section 5 below.

\section{Quantum gravity}

\subsection{The difficulties to match general relativity with quantum theory}

Einstein eq.$\left( \ref{Einst}\right) $ is meaningful in a classical
framework, but it does not fit in quantum theory. In fact the Einstein
tensor of the left side of eq.$\left( \ref{Einst}\right) $ is a classical
(c-number) quantity, but in quantum theory the stress-energy tensor of the
right side should be an observable which in the canonical (HS) formalism is
represented in terms of operators on Hilbert space. Solutions to the problem
might be either substituting a c-number tensor for the stress-energy
operator in the right side or to substitute operators for the components of
the Einstein tensor in the left side. However none of these solutions is
good. An approximation to the former solution might be achieved substituting
the expectation number of the stress-energy operator $\hat{T}_{\mu \nu },$
in the appropriate state $\mid \psi \rangle ,$ for the operator itself. That
is the following would be substituted for eq.$\left( \ref{Einst}\right) $ 
\begin{equation}
G_{\mu \nu }=-8\pi G\left\langle \psi \left| \hat{T}_{\mu \nu
}^{matt}\right| \psi \right\rangle .  \label{Einst2}
\end{equation}
This equation has been used with success in some cases, but it is just a
semiclassical approximation which ignores quantum fluctuations.

The alternative solution would be to promote the metric tensor elements, $%
g_{\mu \nu },$ to be operators, say $\hat{g}_{\mu \nu },$ and hence to get a
quantum operator form of the Einstein tensor $\hat{G}_{\mu \nu }.$ However
there is an ambiguity in passing from $\hat{g}_{\mu \nu }$ to $\hat{G}_{\mu
\nu }$ because the operators $\hat{g}_{\mu \nu }$ would not commute with
each other and with their derivatives, in general. For these reasons people
have attempted to find a theory that unifies quantum theory with general
relativity (GR) via other approaches, a programme known as ``quantum
gravity'' \cite{Rovelli}.

Actually the main motivation for quantum gravity is the attempt to deal with
the problem of singularities in spacetime. Singularities are predicted at
the center of collapsed astrophysical bodies (black holes) and also in the
very early universe. Near a singularity both quantum and ``gravitational''
(i.e. general relativistic) effects are equaly relevant. This is in contrast
with what happens far from singularities. Indeed quantum effects are
relevant in laboratories where gravity may be neglected, and gravity is
relevant in astrophysics where quantum effects are negligible (maybe with
some exceptions, see \cite{Entropy}). In both these regimes quantization of
gravity is not needed. I shall comment on black holes in section 4.3 below.
For the universe there are strong arguments for the assumption that the
universe started about 1.4$\times 10^{10}$ years ago and there was a period
with very strong mass density where both quantum and gravitational effects
were important. The study of early universe is out of the scope of this
article.

\subsection{The search for quantum gravity}

A usual approach to quantum gravity has been to reinterpret general
relativity as a ``field theory of gravity'', which might be quantized like
other relativistic fields. A justification for treating GR as a field theory
has been that there is an alternative road to Einstein eq.$\left( \ref{Einst}%
\right) ,$ not starting from the equivalence principle (i.e. the equivalence
between gravity and acceleration), but deriving GR as a gauge theory whose
associated quantum particle, the graviton, is massless with spin 2. The
meaning of Einstein eq.$\left( \ref{Einst}\right) $ as a relation between
matter and spacetime is usually maintained, whence spacetime itself \emph{%
is taken both }as a quantum field and the ground for all fields, which to me
looks bizarre.

A canonical quantization procedure in analogy to other fields, e.g.
electromagnetic, is not possible due to the nonlinear character of GR. Other
quantization methods have led to theories that are not renormalizable.
Consequently several different routes have been devised without complete
success till now. For instance string theory, loop quantum gravity,
noncommutative geometry and others \cite{Rovelli}.

In section 2.4 I have criticized the opinion that general relativity may be
seen as a field theory of gravity. Therefore I propose a different approach
for the quantization of GR. I support the view that eq.$\left( \ref{Einst}%
\right) $ is just the relation between the curvature of spacetime and the
energy and momentum of the true fields, say those of elementary particle
physics. Thus the unification of quantum theory with general relativity
should consist of the studying quantum fields in curved spacetime, and
attributing to spacetime properties induced by the fact that actual fields
are quantized. In the approach via WW these properties consists essentially
of the existence of vacuum fields, ZPF. Then the stochastic character of the
ZPF would lead to the necessity of assuming a stochastic character to the
curvature of spacetime. That is to \emph{study spacetime} via a
probability distribution of (classical) metrics $g_{\mu \nu }$, each one
giving rise to a different Einstein tensor $G_{\mu \nu }.$ The probability
distribution of metrics should fit in the distribution of energy- momentum
tensors via eq.$\left( \ref{Einst}\right) .$ Thus I conclude that spacetime
has fluctuations at all scales. In any case we should neither treat
spacetime as a field nor gravity as a force (see section 2.3) and maintain
Einstein eq.$\left( \ref{Einst}\right) $ as a valid relation between matter
and spacetime, both treated as random.

\subsection{The problem of singularities. Black holes}

The current opinion is that many stars may collapse after some period of
cooling and/or contraction. The collapse gives rise to a Schwarzshild
singularity in collapsed stars (black holes) \cite{Camenzind}. It is
remarkable that the possible existence of actual, physical, singularities
was rejected by several celebrated authors, including Einstein \cite{Einst39}%
, \cite{Pais}. However a number of theoretical studies have led to accept
that the collapse of spherical compact objects with a high ratio mass by
radius, that is $M/R{>}2G$, is unavoidable. After an early work by
Oppenheimer et al. \cite{OV},\cite{OS} this opinion was strongly advocated
by J. A. Wheeler et al. \cite{Wheeler} around 1960, and it has been
allegedly supported by observations in the years elapsed from that date. The
amount of work on black holes carried out over the last 70 years has been
enormous. Hence it is now current wisdom, although there is still some
controversy about the subject, see \cite{Bambi}. I hope that the subject
will be clarified in the future. The possible singularity associated to the
big bang will not be discussed in this article.

\section{The motion of non-relativistic quantum particles}

The concept of particle in relativistic motion is inappropriate in quantum
theory. (Here relativity refers to the special theory). The reason is that
in the relativistic domain it is necessary to deal with the creation and
annihilation of particles, which is studied by relativistic quantum field
theory. According to the interpretation of this article, Bose fields are
wavelike, the particle behaviour being an effect of the ZPF, as will be
discussed in detail section 4. For Fermi fields there is no formulation
similar to WW, able to provide a clear realistic picture. Then the following
will be devoted to a realistic interpretation of the mechanics of particles
in non-relativistic motion, that is with velocity much smaller than the
speed of light.

As discussed in section 3.3 WW does not provide a realistic picture for
particles because they are always dressed with quantum fields, which
modifies strongly the motion. In the following I propose an interpretation
resting on the formulation of quantum mechanics via path-integrals,
introduced by Feynman in 1948. The application to non-relativistic quantum
mechanics is studied in a book by Feynman and Hibbs \cite{FeynmanHibbs}. In
the following I review briefly the realistic interpretation presented
elsewhere \cite{Entropy2}.

\subsection{Path-integral formulation of quantum mechanics}

Feynman proposed to calculate the amplitude that a particle placed at
position x$_{0}$ at time t=0 reaches the position x at time t, as follows
(in one dimension)

\begin{equation}
A(x_{0},0\mid x,t)=\int dx_{1}...\int dx_{n-1}A(x_{0},0\mid
x_{1},t_{1})...A(x_{n-1},t_{n-1}\mid x,t).  \label{Feyn6}
\end{equation}
The set of positions $\left\{ x_{0},x_{1},...x\right\} $ defines a
(discrete) path, whence eq.$\left( \ref{Feyn6}\right) $ is an integral of
discrete paths. From eq.$\left( \ref{Feyn6}\right) $ we may obtain the
corresponding probability, that is 
\begin{equation}
P(x)\propto \left| A(x_{0},0\mid x,t)\right| ^{2}.  \label{Feyn}
\end{equation}
The time intervals may be chosen identical, that is $t_{j+1}-t_{j}=%
\varepsilon ,$ with $\varepsilon $ as small as desired. In the limit $%
\varepsilon \rightarrow 0,$ $A(x_{0},0\mid x,t)$ becomes an integral of
continuous path amplitudes$.$

In the case of one-dimensional motion in a potential $V(x)$ the partial
amplitudes are defined as follows 
\begin{eqnarray}
A\left( x_{j-1},t_{j-1}\mid x_{j},t_{j}\right) &=&\sqrt{\frac{m}{2\pi i 
\rlap{\protect\rule[1.1ex]{.325em}{.1ex}}h%
\varepsilon }}\exp (\frac{i\varepsilon }{
\rlap{\protect\rule[1.1ex]{.325em}{.1ex}}h%
}L_{j}),\smallskip \smallskip  \label{Feyn4} \\
L_{j} &\equiv &\frac{1}{2}m\left( \frac{x_{j}-x_{j-1}}{\varepsilon }\right)
^{2}-\frac{1}{2}\left[ V\left( x_{j-1}\right) +V\left( x_{j}\right) \right] 
\nonumber
\end{eqnarray}
where $m$ is the mass of the particle. (This expression differs from the
original one of Feynman\cite{FeynmanHibbs} because I have substituted $\frac{%
1}{2}\left[ V\left( x_{j-1}\right) +V\left( x_{j}\right) \right] $ for $%
V\left[ \left( x_{j-1}+x_{j}\right) /2\right] $ for later convenience. Both
formulations agree in the limit $\varepsilon \rightarrow 0).$

The amplitude $A(x_{0},0\mid x,t)$ is named the ``propagator'' of the
wavefunction $\psi \left( x,t\right) $, it allows getting the wavefunction
at time $t$ from the wavefunction at time $0$, that is 
\begin{equation}
\psi \left( x,t\right) =\int dx_{0}\psi \left( x_{0},t\right) A(x_{0},0\mid
x,t).  \label{F9}
\end{equation}
Hence the propagator $A(x_{0},0\mid x,t)$ fulfils the Schr\"{o}dinger
equation with the initial condition 
\[
A(x_{0},0\mid x,0)=\delta \left( x-x_{0}\right) , 
\]
where $\delta \left( x\right) $ is Dirac\'{}s delta. Thus the propagator is
the Green\'{}s function of the Schr\"{o}dinger equation.

The path integrals formulation may be generalized to 3 dimensions, to
many-particles and also to relativistic field theory. It has an extremely
important role in modern theoretical physics, both because it is well
adapted to derive general properties, e.g. symmetries, and due to the
relevance for actual calculations, it being the seed of Feynman graphs in
covariant perturbation theory \cite{ZinnJustin}. Dealing with formal and
calculational aspects lies out of the scope of this section, which is
devoted to the physical interpretation of the Feynman formalism in
(non-relativistic) quantum mechanics.

\subsection{Transition probability as a sum of paths probabilities}

In the following I present a formulation for the motion of a quantum
particle in terms of probabilities (rather than amplitudes!) of paths, with
the condition that the transition probability agrees with the square modulus
of the Feynman amplitude eq.$\left( \ref{Feyn}\right) ,$ that is 
\[
P(x_{0},0\rightarrow x,t)=\left| A(x_{0},0\mid x,t)\right| ^{2}. 
\]
If we take the (continuous)\ set of paths as discrete for the sake of
clarity, and we generalize to 3 dimensions, our aim is to get the transition
probability as a sum of probabilities of paths, that is 
\begin{equation}
P(\mathbf{r}_{a},t_{a}\rightarrow \mathbf{r}_{b},t_{b})=\sum_{k}W_{k}(%
\mathbf{r}_{a},t_{a}\rightarrow \mathbf{r}_{b},t_{b}).  \label{3.2}
\end{equation}
Here every value of the index $k$ corresponds to a possible path of the
particle with end points $(\mathbf{r}_{a},t_{a})$ and $(\mathbf{r}%
_{b},t_{b}) $. The problem is to find ``weights'' $W_{k}$ which could be
interpreted as probabilities, in order to provide an intuitive picture of
the quantum evolution as a random motion of particles. In the following I
propose a method to get the said weights. In some cases to be studied below,
the weights are non-negative definite and therefore may be interpreted as
probabilities, in other cases the formalism should be slightly modified in
order to get positivity.

I shall start from the 3D generalization of the amplitude eq.$\left( \ref
{Feyn4}\right) $, that is 
\begin{eqnarray}
A\left( \mathbf{x}_{a},t_{a}\rightarrow \mathbf{x}_{b}\mathbf{,}t_{b}\right)
&=&\lim_{\varepsilon \rightarrow 0}\left( \frac{m}{2\pi i 
\rlap{\protect\rule[1.1ex]{.325em}{.1ex}}h%
\varepsilon }\right) ^{3n/2}\int d\mathbf{x}_{n-1}...\int d\mathbf{x}_{1}
\label{2.9} \\
&&\prod_{j=1}^{n}\exp \left\{ \frac{im}{2
\rlap{\protect\rule[1.1ex]{.325em}{.1ex}}h%
\varepsilon }\left| \mathbf{x}_{j}-\mathbf{x}_{j-1}\right| ^{2}-\frac{%
i\varepsilon }{2
\rlap{\protect\rule[1.1ex]{.325em}{.1ex}}h%
}\left[ V\left( \mathbf{x}_{j-1}\right) +V\left( \mathbf{x}_{j}\right)
\right] \right\} ,  \nonumber
\end{eqnarray}
where $\varepsilon \equiv t_{j}-t_{j-1},$ is independent of $j$ and $\mathbf{%
x}_{0}\equiv \mathbf{x}_{a},\mathbf{x}_{n}\equiv \mathbf{x}_{b}.$ The limit $%
\varepsilon \rightarrow 0$ should be understood with $n\rightarrow \infty $
fulfilling 
\begin{equation}
\lim_{\varepsilon \rightarrow 0}\left( n\varepsilon \right) =t_{b}-t_{a}.
\label{nepsilon}
\end{equation}
Actually the integrals involved are not convergent, therefore an appropriate
regularization is implicit.

The transition probability is the square modulus of the transition
amplitude, which becomes \cite{Entropy2} 
\begin{eqnarray}
P(\mathbf{r}_{a},t_{a} &\rightarrow &\mathbf{r}_{b},t_{b})=A\left( \mathbf{x}%
_{a},t_{a}\rightarrow \mathbf{x}_{b}\mathbf{,}t_{b}\right) A^{*}\left( 
\mathbf{y}_{a},t_{a}\rightarrow \mathbf{y}_{b}\mathbf{,}t_{b}\right) 
\nonumber \\
&=&\lim_{\varepsilon \rightarrow 0}\left( \frac{m}{2\pi 
\rlap{\protect\rule[1.1ex]{.325em}{.1ex}}h%
\varepsilon }\right) ^{3n}\int d\mathbf{x}_{n-1}...\int d\mathbf{x}_{1}\int d%
\mathbf{y}_{n-1}...\int d\mathbf{y}_{1} \\
&&\times \prod_{j=1}^{n}\exp \left\{ \frac{im}{2\varepsilon 
\rlap{\protect\rule[1.1ex]{.325em}{.1ex}}h%
}\left[ \left| \mathbf{y}_{j}-\mathbf{y}_{j-1}\right| ^{2}-\left| \mathbf{x}%
_{j}-\mathbf{x}_{j-1}\right| ^{2}\right] \right\}  \nonumber \\
&&\times \prod_{j=1}^{n}\exp \left\{ -\frac{i\varepsilon }{2 
\rlap{\protect\rule[1.1ex]{.325em}{.1ex}}h%
}\left[ V(\mathbf{x}_{j})-V(\mathbf{y}_{j})\right] \right\} ,  \label{3.1}
\end{eqnarray}
where I identified $\mathbf{x}_{a}=\mathbf{y}_{a}=\mathbf{r}_{a},$ $\mathbf{x%
}_{b}=\mathbf{y}_{b}=\mathbf{r}_{b}$ and reordered the integrals. I have
represented vectors with bold face letters, so that $\int d\mathbf{x}_{j}$ , 
$\int d\mathbf{y}_{j}$ are triple integrals over the whole 3D space and I
have included the parameters $m$ and $
\rlap{\protect\rule[1.1ex]{.325em}{.1ex}}h%
$ following Feynman \cite{FeynmanHibbs}. With that choice the quantity $P(%
\mathbf{r}_{a},t_{a}\rightarrow \mathbf{r}_{b},t_{b})$ has dimensions of
probability per square volume. Then the probability that a particle is in
some finite volume $B$ at time $t_{b}$ conditional to be in another finite
volume $A$ at an earlier time $t_{a}$ will be 
\[
P\left( A\rightarrow B\right) =\int_{\mathbf{r}_{a}\in A}d\mathbf{r}%
_{a}\int_{\mathbf{r}_{b}\in B}d\mathbf{r}_{b}P(\mathbf{r}_{a},t_{a}%
\rightarrow \mathbf{r}_{b},t_{b}). 
\]

In order to proceed I shall make a change of variables, that is 
\begin{equation}
\mathbf{r}_{j}=\frac{1}{2}\left( \mathbf{x}_{j}+\mathbf{y}_{j}\right) ,%
\mathbf{u}_{j}\mathbf{=x}_{j}\mathbf{-y}_{j},0\leq j\leq n.  \label{newvar}
\end{equation}
Hence eq.$\left( \ref{3.1}\right) $ becomes, reordering the exponentials, 
\begin{eqnarray}
P(\mathbf{r}_{a},t_{a} &\rightarrow &\mathbf{r}_{b},t_{b})=\lim_{\varepsilon
\rightarrow 0}\left( \frac{m}{2\pi \varepsilon 
\rlap{\protect\rule[1.1ex]{.325em}{.1ex}}h%
}\right) ^{3n}\int d\mathbf{r}_{n-1}...\int d\mathbf{r}_{1}\int d\mathbf{u}%
_{n-1}...\int d\mathbf{u}_{1}  \nonumber \\
&&\times \prod_{j=1}^{n-1}\exp \left[ -\frac{im}{\varepsilon 
\rlap{\protect\rule[1.1ex]{.325em}{.1ex}}h%
}\mathbf{u}_{j}\cdot \left( \mathbf{r}_{j-1}\mathbf{-}2\mathbf{r}_{j}\mathbf{%
+r}_{j+1}\right) \right]  \nonumber \\
&&\times \prod_{j=1}^{n-1}\exp \left\{ \frac{i\varepsilon }{2 
\rlap{\protect\rule[1.1ex]{.325em}{.1ex}}h%
}\left[ V(\mathbf{r}_{j}-\frac{1}{2}\mathbf{u}_{j})-V(\mathbf{r}_{j}+\frac{1%
}{2}\mathbf{u}_{j})\right] \right\} ,\smallskip  \label{3.3}
\end{eqnarray}
where $\mathbf{r}_{0}=\mathbf{r}_{a}$, $\mathbf{r}_{n}=\mathbf{r}_{b}$ and $%
\mathbf{u}_{0}=\mathbf{u}_{n}=0$, whence the probability may be written 
\begin{equation}
P(\mathbf{r}_{a},t_{a}\rightarrow \mathbf{r}_{b},t_{b})=\lim_{\varepsilon
\rightarrow 0}\left( \frac{m}{2\pi 
\rlap{\protect\rule[1.1ex]{.325em}{.1ex}}h%
}\right) ^{3}\varepsilon ^{-3n}\int d\mathbf{r}_{n-1}...\int d\mathbf{r}%
_{1}\prod_{j=1}^{n-1}Q_{j},  \label{3.4}
\end{equation}
where 
\begin{eqnarray}
Q_{j} &\equiv &\left( \frac{m}{2\pi 
\rlap{\protect\rule[1.1ex]{.325em}{.1ex}}h%
}\right) ^{3}\int d\mathbf{u}\exp \left( -i\frac{m}{%
\rlap{\protect\rule[1.1ex]{.325em}{.1ex}}h%
}\mathbf{u\cdot s}_{j}\right)  \nonumber \\
&&\times \exp \left\{ \frac{i\varepsilon }{2
\rlap{\protect\rule[1.1ex]{.325em}{.1ex}}h%
}\left[ V(\mathbf{r}_{j}-\frac{1}{2}\mathbf{u})-V(\mathbf{r}_{j}+\frac{1}{2}%
\mathbf{u})\right] \right\} ,  \label{7.15}
\end{eqnarray}
with 
\begin{equation}
\mathbf{s}_{j}\equiv \frac{\mathbf{r}_{j-1}\mathbf{-}2\mathbf{r}_{j}\mathbf{%
+r}_{j+1}}{\varepsilon }=\frac{\mathbf{r}_{j+1}\mathbf{-r}_{j}}{\varepsilon }%
-\frac{\mathbf{r}_{j}\mathbf{-r}_{j-1}}{\varepsilon }\equiv \mathbf{v}_{j}%
\mathbf{-v}_{j-1},j=1,2,...n-1.  \label{7.14b}
\end{equation}
The quantity $\mathbf{s}_{j}$ has the physical meaning of velocity change at
time $t$ and the ratio $\mathbf{s}_{j}/\varepsilon $ might be interpreted as
an acceleration. However the limit $\varepsilon \rightarrow 0$ may not
exist, that is the instantaneous velocity and acceleration are not well
defined in general. I will return to that point in section 4.2 below.

Performing the integrals in $\mathbf{u}_{j}$ is not possible without a
knowledge of the potential, $V\left( \mathbf{r}\right) ,$ but to lowest
order in $
\rlap{\protect\rule[1.1ex]{.325em}{.1ex}}h%
$ the integrals are simple. In fact approximating $V\left( \mathbf{r}_{j}\pm 
\mathbf{u}_{j}/2\right) $ to first order in $\mathbf{u}_{j}$ in the second
exponent of eq.$\left( \ref{7.15}\right) $ and then integrating with respect
to $\mathbf{u,}$ I get 
\begin{eqnarray}
P(\mathbf{r}_{a},t_{a} &\rightarrow &\mathbf{r}_{b},t_{b})=\lim_{\varepsilon
\rightarrow 0}\left( \frac{m}{2\pi 
\rlap{\protect\rule[1.1ex]{.325em}{.1ex}}h%
\varepsilon }\right) ^{3\left( n+1\right) /2}\int d\mathbf{r}_{1}...\int d%
\mathbf{r}_{n-1}  \label{F0} \\
&&\times \prod_{j=1}^{n-1}\delta ^{3}\left( \mathbf{s}_{j}+\frac{%
2\varepsilon }{m}\mathbf{\nabla }V(\mathbf{r}_{j})\right) +O\left( 
\rlap{\protect\rule[1.1ex]{.325em}{.1ex}}h%
^{3}\right) ,  \nonumber
\end{eqnarray}
This corresponds to a motion fulfilling at every time 
\begin{equation}
m\frac{\mathbf{r}_{j-1}-2\mathbf{r}_{j}+\mathbf{r}_{j+1}}{\varepsilon ^{2}}=-%
\mathbf{\nabla }V(\mathbf{r}_{j}),  \label{F2}
\end{equation}
that is the (discretized) classical equation of motion. Thus eq.$\left( \ref
{F0}\right) $ provides the classical limit of quantum mechanics when $
\rlap{\protect\rule[1.1ex]{.325em}{.1ex}}h%
\rightarrow 0.$

It is interesting that the classical motion, eq.$\left( \ref{F2}\right) ,$
is obtained without any approximation when the potential, $V(\mathbf{r}%
_{j}), $ is at most quadratic in the coordinates because in this case eq.$%
\left( \ref{F2}\right) $ is exact (no term $O\left( 
\rlap{\protect\rule[1.1ex]{.325em}{.1ex}}h%
^{3}\right) $ appears). This might be interpreted saying that in ``linear
problems the quantum particle follows the classical path''. The typical
example is the harmonic oscillator. This is the reason why quantum mechanics
of linear systems looks semiclassical. In this case all quantum effects
derive from the fact that the initial wave function cannot be localized in a
too small region due to the Heisenberg uncertainty principle, a constraint
which does not appear in Feynman\'{}s path integrals formalism, and should
be put as an additional constraint. In contrast it does appear in the
canonical Hilbert space formalism, where Heisenberg uncertainty relations
are a consequence of the commutation rules.

\subsection{Path weights in terms of the Fourier transform of the potential}

It is convenient to perform a change leading to a more simple description of
the transition probability, but equivalent to eq.$\left( \ref{3.3}\right) $
in the limit $\varepsilon \rightarrow 0,n\rightarrow \infty .$ After some
algebra we get the following transition probability \cite{Entropy2}

\begin{eqnarray}
P(\mathbf{r}_{a},t_{a} &\rightarrow &\mathbf{r}_{b},t_{b})=\lim_{\varepsilon
\rightarrow 0}\left( \frac{m}{2\pi 
\rlap{\protect\rule[1.1ex]{.325em}{.1ex}}h%
}\right) ^{3}\varepsilon ^{-3n}\int d\mathbf{r}_{n-1}...\int d\mathbf{r}%
_{1}\prod_{j=1}^{n-1}Q_{j},Q_{j}=D_{j}+\varepsilon F_{j}  \nonumber \\
D_{j} &\equiv &\delta ^{3}\left( \mathbf{s}_{j}\right) ,F_{j}\equiv -\frac{%
m^{3}}{2
\rlap{\protect\rule[1.1ex]{.325em}{.1ex}}h%
^{4}}\operatorname{Im}\left[ \tilde{V}\left( \frac{2m\mathbf{s}_{j}}{%
\rlap{\protect\rule[1.1ex]{.325em}{.1ex}}h%
}\right) \exp \left( -\frac{2im\mathbf{s}_{j}\mathbf{\cdot r}_{j}}{%
\rlap{\protect\rule[1.1ex]{.325em}{.1ex}}h%
}\right) \right] ,  \label{77}
\end{eqnarray}
where $\mathbf{r}_{o}=\mathbf{r}_{a},\mathbf{r}_{n}=\mathbf{r}_{b},$ and $%
\mathbf{s}_{j}$ is the change of velocity at time $t_{j},$ see eq.$\left( 
\ref{7.14b}\right) .$ The Fourier transform is here defined as follows 
\begin{equation}
\tilde{V}\left( \mathbf{w}\right) \equiv \int d\mathbf{x}\exp \left( i%
\mathbf{w.x}\right) V\left( \mathbf{x}\right) ,  \label{Fourier}
\end{equation}
$\mathbf{w}$ and $\mathbf{x}$ being 3D vectors. Calculating the transition
probability $P(\mathbf{r}_{a},t_{a}\rightarrow \mathbf{r}_{b},t_{b})$ is
involved because the changes of the velocity and the positions are related
via eqs.$\left( \ref{7.14b}\right) .$

\subsection{Realistic interpretation}

Two interesting questions are whether the paths involved in eq.$\left( \ref
{77}\right) $ are continuous and whether the quantities $Q_{j}$ are positive
(or zero). The answers to both questions are affirmative \cite{Entropy2}.
However the quantities $\mathbf{s}_{j}$, which would represent the
instantaneous acceleration in the limit $\varepsilon \rightarrow 0$, are not
well defined. That is the functions $\mathbf{r}\left( t\right) $ are
continuous but only once derivable.

Eq.$\left( \ref{77}\right) $ afford a formulation of the quantum motion of
particles in the form of a probability distribution of possible paths from
the initial position $\mathbf{r}_{a}$ at time $t_{a}$ to a final position $%
\mathbf{r}_{b}$ at time $t_{b}.$ The motion has a random character, although
quite different from the most popular Brownian motion. It is interesting
that the probability of every path depends on the Fourier transform $\tilde{V%
}\left( 2s_{j}\right) $ of the potential, $V\left( \mathbf{r}\right) ,$
whence the effect of $V\left( \mathbf{r}\right) $ on the motion of the
particle is non-local. A plausible explanation for this fact is that the
particle motion is influenced by the fluctuating spacetime curvature
discussed in section 3.6, and also by the interaction with the vacuum fields
discussed in section 3.4.

It is remarkable that these influences, rather involved, give rise to a
relatively simple action via de Fourier transform of the external force
(here treated as deriving of a potential $V\left( \mathbf{r}\right) $). This
may be put in a different form, namely the question: how the quantum effects
may be taken into account via the rather simple mathematical formalism of
Hilbert spaces?. Indeed we have shown that Feynman path integral formalism
for quantum particles is equivalent to Schr\"{o}dinger formulation, and this
is equivalent to the canonical HS formalism.

\subsection{Scattering experiments. Born approximation}

An application of the formalism here proposed is the study of scattering of
a particle by a potential. In those experiments a source emits particles,
all of them with velocity $\mathbf{v}_{a}$. A fraction of the particles
cross a target region where they experience the force due to the potential $%
V\left( \mathbf{r}\right) .$ Then the particles emerge from the target with
velocities $\left\{ \mathbf{v}_{b}\right\} $ different from the initial one $%
\mathbf{v}_{a}$ and eventually arrive at a detector. The target is in
practice small (microscopic) while the distances from the target to either
the source or the detector are both large (macroscopic). In the formalism of
this article I assume that the particles are small (or pointlike)
corpuscles. No waves appear.

The quantity of interest in scattering experiments is the differential cross
section, $\sigma \left( \theta ,\phi \right) $. It is proportional to the
number of particles per unit solid angle that leave the target with a
velocity $\mathbf{v}_{b}$ in the direction determined by the angles $\left(
\theta ,\phi \right) $. In our formalism we may write 
\begin{equation}
\sigma \left( \theta ,\phi \right) \propto \int d\mathbf{r}_{a}\rho \left( 
\mathbf{r}_{a}\right) \int v_{b}^{2}dv_{b}P_{v}\left( \mathbf{r}%
_{a}\rightarrow \mathbf{v}_{b}\right) ,  \label{7.1}
\end{equation}
where $P_{v}\left( \mathbf{v}_{b}\right) $ is the probability that a
particle emerging from the point $\mathbf{r}_{a}$ of the source
with velocity $\mathbf{v}_{a}$ reaches the velocity $\mathbf{v}_{b}$ after
crossing the target. The integral with respect to the modulus of $\mathbf{v}%
_{b}$ takes into account that only the direction of $\mathbf{v}_{a}$
matters, not the modulus. The triple integral with respect to $\mathbf{r}%
_{a} $ is necesary in order to sum over all possible initial positions of
the particle in the source. However the cross section should be independent
of the density $\rho \left( \mathbf{r}_{a}\right) $ of particles. Then I
shall assume that the velocity is in the direction of the Z axis and the
density $\rho $ is homogeneous in a slab with limits $\alpha \leq z\leq
\beta .$ Then the following should be substituted for eq.$\left( \ref{7.1}%
\right) $ 
\begin{equation}
\sigma \left( \theta ,\phi \right) \propto \int_{-\infty }^{\infty
}dx_{a}\int_{-\infty }^{\infty }dy_{a}\int v_{b}^{2}dv_{b}P_{v}\left( 
\mathbf{r}_{a}\rightarrow \mathbf{v}_{b}\right) .  \label{7.2}
\end{equation}
Calculating exactly $P_{v}\left( \mathbf{v}_{b}\right) $ is involved, but it
is relatively simple in the Born approximation.

In our approach the Born approximation consists of writing the product $%
\prod_{j=1}^{n-1}Q_{j}$ as an expansion in powers of $\varepsilon F$ taking
eq.$\left( \ref{77}\right) $ into account and truncating the expansion to
second order, that is 
\begin{eqnarray}
\prod_{j=1}^{n-1}Q_{j} &=&\prod_{j=1}^{n-1}(D_{j}+\varepsilon F_{j})\simeq
\prod_{j=1}^{n-1}D_{j}+\sum_{k}\left\{ \prod_{j=1}^{k-1}D_{j}\right\}
\varepsilon F_{k}\left\{ \prod_{l=k+1}^{n-1}D_{l}\right\}  \nonumber \\
&&+\sum_{ki}\left\{ \prod_{j=1}^{k-1}D_{j}\right\} \varepsilon F_{k}\left\{
\prod_{l=k+1}^{i-1}D_{l}\right\} \varepsilon F_{i}\left\{
\prod_{r=l+1}^{n-1}D_{r}\right\} .  \label{Born}
\end{eqnarray}
It can be shown that the sum $\sum_{k}\varepsilon F_{k}$ consists of $n$
terms whence it remains finite in the limit $\varepsilon \rightarrow 0,$ see
eq.$\left( \ref{nepsilon}\right) .$ Therefore the small parameter in the
expansion is actually the potential $V$, see eq.$\left( \ref{77}\right) $,
as is typical in the Born approximation. Products like $%
\prod_{j=1}^{k-1}D_{j}$ correspond to motion in straight line and constant
velocity from the time $t_{j}$ to the time $t_{k-1}$. Terms like $F_{k}$
give the probabilities of the possible changes of velocity $s_{k}$ at time $%
t_{k},$ see $\left( \ref{77}\right) .$

The first two terms of eq.$\left( \ref{Born}\right) $ do not contribute to
the cross section, whence the third term is dominant in the expansion. That
term leads to the well known result of Born cross section \cite{Entropy2}.
That is, in terms of the initial and final wavevectors associated to the
particle, $\mathbf{k}_{a}$ and $\mathbf{k}_{b}$ respectively, we get 
\[
\sigma \left( \theta ,\phi \right) =\frac{1}{16\pi ^{2}}\left| \int d\mathbf{%
x}\exp \left[ -i\mathbf{x\cdot }\left( \mathbf{k}_{b}-\mathbf{k}_{a}\right)
\right] V\left( \mathbf{x}\right) \right| ^{2},\mathbf{k=}\frac{m\mathbf{v}}{%
\rlap{\protect\rule[1.1ex]{.325em}{.1ex}}h%
}. 
\]

\subsection{Interference experiments with particles}

The wave behaviour of electrons, proposed by L. de Broglie in 1923, led soon
to experiments proving their interference. Later on similar experiments have
been performed with neutrons, atoms and even molecules. The results may be
explained with the formalism presented in this section.

Indeed Born's approximation allows calculating the result of a simple
interference experiment. In fact let us consider a particle with initial
velocity $\mathbf{v}_{0}=\left( 0,0,v_{0}\right) ,$ which is moving in the $%
Z $ direction and eventually arrives at a region with the potential 
\begin{equation}
V\left( \mathbf{r}\right) =C\left\{ \exp \left[ -\lambda \left( \mathbf{r+a}%
\right) ^{2}\right] +\exp \left[ -\lambda \left( \mathbf{r-a}\right)
^{2}\right] \right\} ,\mathbf{a\equiv }\left( a,0,0\right) ,  \label{17}
\end{equation}
which is a model for a screen with two holes, chosen for an easy
calculation. Obtaining the cross section via Born's approximation is not
difficult using eq.$\left( \ref{Born}\right) .$ We get 
\[
\sigma \propto \exp \left[ -\frac{\mathbf{v}^{2}+\mathbf{v}%
_{0}^{2}-2v_{0}v_{z}}{2\lambda }\right] \cos ^{2}\left( av_{x}\right) , 
\]
where $\mathbf{v=}\left( v_{x},v_{y},v_{z}\right) $ is the final velocity$.$
Hence the cross section becomes, taking the conservation of energy into
account, 
\begin{equation}
\sigma =C^{2}\frac{\pi }{\lambda ^{3}}\exp \left[ -\frac{2v_{0}^{2}\sin
^{2}\theta }{\lambda }\right] \cos ^{2}\left( av_{0}\sin \theta \cos \phi
\right) .  \label{18}
\end{equation}
Assuming that particle detections are observed as spots produced in a screen
placed parallel to the $XY$ plane, we would observe typical interference
fringes with a decreasing intensity in both directions $X$ and $Y$ and a
maximum at $x=y=0$.

The point of this calculation is that a wave behaviour of particles is
absent, particles remaind during the interference experiments. The wave
behaviour in the interference is an effect of the non-local action of the
potential. However it is plausible to assume that the said action is
mediated by some ``hidden'' waves, which fits in the assumption that quantum
particles are not simple objects, but they are ''dressed'' with waves at a
difference with the classical ones. In fact I propose that the mentioned
waves may be just some combination of the fluctuating spacetime and the ZPF
of all quantum fields, in particular the electromagnetic one, but not only
it.

\subsection{Discussion}

I have shown that in non-relativistic quantum mechanics (without spin) it is
possible to picture the transition probability in terms of particle paths. A
path may be defined by the positions $\left\{ \mathbf{r}_{a}\equiv \mathbf{r}%
_{0},...\mathbf{r}_{j},...\mathbf{r}_{b}\equiv \mathbf{r}_{n}\right\} $ at
times $t_{a,}...t_{j}\equiv t_{a}+j\varepsilon ,...t_{b}$ or, what is
equivalent, the initial and final positions plus the velocity changes $%
\left\{ \mathbf{s}_{j}/\varepsilon \right\} $ at times $t_{j}.$ Eventually
we should consider the limit $n\rightarrow \infty $ with $n\varepsilon
=t_{b}-t_{a}.$

Is summary the formalism suggests an intuitive picture of non-relativistic
quantum mechanics in terms of \emph{probabilities of the possible paths of}
\emph{particles}. The particle\'{}s motion is represented by a stochastic
process such that there is a random change of velocity at every time $t_{j}$%
, with a probability depending on \emph{the potential over a large region}
around the position of the particle (indeed it derives from the Fourier
transform of the potential, see eq.$\left( \ref{77}\right) $). The wave
behaviour, e. g. in experiments of atom interference, may be interpreted
assuming that the motion of the particles is governed by a law (different
from Newton\'{}s) where the ``acceleration'' depends on the potential on a
whole spatial region, at a difference with the local action of classical
dynamics. I have dealt with a single particle, but the generalization to $N$
interacting particles is straightforward, except for the possible effects of
quantum statistics.

With this interpretation the interference experiments with particles (e.g.
atoms) might be explained without assuming that those particles possess a
wave nature or that they may cross two distant slits at the same time. But I
stress again that we remain at the level of non-relativistic quantum
mechanics. I do not claim that a similar interpretation may be extended to
relativistic quantum field theory when spin plays a role, or even to atoms
or molecules when (Bose or Fermi) statistics is relevant.

\section{Wave-particle duality}

\subsection{Realistic interpretation of photon effects via the ZPF}

The difficulty to reconcile the wave and particle behaviour of light is a
big obstacle for a classical-like interpretation of quantum phenomena. In
particular the celebrated experiment of anticorrelation and recombination of
light after a beam splitter \cite{Grangier} (see below) has been used in
popular books in order to argue against the possibility of getting pictures
of reality for the quantum phenomena, e.g. \cite{Rovelli1}.

The origin of the wave-particle duality goes back to Einstein\'{}s proposal
that light consists of particles, later called photons. Hence he derived
successfully the laws of the photoelectric effect (an achievement which was
mentioned as merit for his Nobel Prize \cite{Pais}). Einstein proposal of
photons was actually unnecessary, because the photoelectric laws may be
derived from the \emph{weaker assumption} that light is absorbed in
discrete amounts of energy $h\nu ,$ which had been proposed (or suggested)
five years earlier by Planck. It is true that Einstein seemed not too happy
with the dual nature of light and attempted a fusion of waves and particles
in later articles\emph{\ }\cite{photons}. Indeed he introduced in 1916 the
concept of radiation needless, as a kind of alternative to ``particles of
light'' (photons).

In the canonical (HS) formalism some reality is ascribed to a zeropoint
field (ZPF) assuming the existence of vacuum fluctuations. These
fluctuations are supposed to consist of ``short lived virtual particles'', a
sentence that is senseless in our realistic interpretation due to the
ambiguous meaning of the word `virtual'. Anyway the ZPF has a great
relevance in the quantitative prediction of measurement results, both in
quantum electrodynamics (QED) and in quantum optics (see e.g. \cite{Milonni}%
). In the WW formalism the predicted EM radiation filling space is
interpreted as a real stochastic field, as discussed in section 3.4. 

In the following I present several phenomena which in principle might be
quantitative interpreted by standard quantum (field) theory, but the
eventual calculation does not provide a picture of reality. Therefore I
shall propose only heuristic qualitative or semiquantitative models in the
followint.

\subsection{Photon detection. Photocounts.}

In order to take account of ``photon detection'', firstly I point out that
the absorption of light in the form of localized spots in a photographic
plate or clicks in a photodetector are not valid arguments for the particle
behaviour of radiation. In fact the former are caused by the granular
(atomic or molecular) nature of the plate, and the photocounts in a detector
derive from the fact that photon counters are manufactured so that they
click whenever the radiation arriving at the detector transfers to it enough
energy, which is compatible with light being waves \cite{Frontiers}, \cite
{FOOP}. It is true that in this case discriminating (weak) light signals,
coming to the detectorom a source, from the assumed (strong) vacuum
radiation (ZPF) looks like searching for a needle in a haystack. However
there are models for photodetection that avoid the problem \cite{Frontiers}, 
\cite{FOOP}. The models take advantage of the fact that the ZPF flux comes
from all directions, it has rotational invariance on the average. Hence we
may assume that the clean effect is nil because the pressure on different
directions cancel out. In contrast the radiation coming from a source is
directional.

Let us now compare the interpretation of the radiation energy measurement in
either the canonical or WW formalisms. In HS the probability distribution of
values got in the measurement of an observable $\hat{M}$ in the state with
density matrix $\hat{\rho},$ may be obtained via the moments, that is the
expectation values of powers of the observable, $\left\langle
M^{n}\right\rangle ,$ as follows 
\begin{equation}
\left\langle M^{n}\right\rangle =Tr\left( \hat{M}^{n}\hat{\rho}\right) .
\label{moments}
\end{equation}
In WW the expectation becomes an integral of the observable written in terms
of the amplitudes, weighted by the probability distribution in the (mixed)
state, see eq.$\left( \ref{expect}\right) $. In the case of the EM field the
most relevant observable is the energy, where $\hat{M}$ becomes the
Hamiltonian operator$.$

As an illustration let us calculate the expectation value of the energy
within WW for the general state $W_{\psi }^{total}$ given in eq.$\left( \ref
{stateWW}\right) $. In order to agree with HS predictions we shall use the
normally ordered Hamiltonian eq.$\left( \ref{HnormalW}\right) .$ For a state 
$\mid \phi \rangle $ of the field the calculation, firstly in HS then in WW,
may be written 
\begin{eqnarray}
\left\langle E\right\rangle &=&\left\langle \phi \left| \sum_{l} 
\rlap{\protect\rule[1.1ex]{.325em}{.1ex}}h%
\omega _{l}\hat{a}_{l}^{\dagger }\hat{a}_{l}\right| \phi \right\rangle 
\nonumber \\
&=&\sum_{l}
\rlap{\protect\rule[1.1ex]{.325em}{.1ex}}h%
\omega _{l}\int W_{\phi }\left( \left\{ a_{l}\right\} \right) \left( \left|
a_{l}\right| ^{2}-\frac{1}{2}\right) d\operatorname{Re}a_{l}d\operatorname{Im}a_{l} 
\nonumber \\
&=&\sum_{l}
\rlap{\protect\rule[1.1ex]{.325em}{.1ex}}h%
\omega _{l}\int \left| a_{l}\right| ^{2}\left( W_{\phi }\left( \left\{
a_{l}\right\} \right) -W_{0}\left( \left\{ a_{l}\right\} \right) \right) d%
\operatorname{Re}a_{l}d\operatorname{Im}a_{l},  \label{photocounts}
\end{eqnarray}
where $W_{\phi }\left( \left\{ a_{l}\right\} \right) $ is the state (Wigner
function) that in WW represents the HS state $\mid \phi \rangle $ and $W_{0}$
the vacuum Wigner function eq.$\left( \ref{1}\right) .$ The result is that
the ZPF does not contribute to detection in agreement with the HS result.
The property may be stated saying that photodetectors are sensitive only to
radiation that exclude the ZPF. For a physical realistic interpretation of
eq.$\left( \ref{photocounts}\right) $ in WW and a more extended discussion
about photodetectors see \cite{FOOP}.

\subsection{ Absorption of light in discrete amounts and the photoelectric
effect}

As is well known, in order to derive his radiation law, Planck introduced in
1900 the hypothesis that absorption and emission of radiation takes place in
``quanta'' of energy $E=
\rlap{\protect\rule[1.1ex]{.325em}{.1ex}}h%
\omega .$ In the following I will show that absorption in discrete amounts
may be explained by the action of the ZPF.

It is plausible that absorption of radiation takes place via resonance with
material oscillators. Thus we may consider a light signal with wavevector $%
\mathbf{k}_{0}$ that arrives at a material having weakly bound electrons
whose motion posseses a relevant component with frequency $c\left| \mathbf{k}%
_{0}\right| $ and it is parallel to $\mathbf{k}_{\mathbf{0}}.$ The ZPF may
be described in terms of plane waves and we are interested in those having
wavevectors $\mathbf{k}$ near $\mathbf{k}_{0}$. From time to time it may
happen that several of these waves have phases close to the incoming signal,
say in a frequency range $\Delta \omega ,$so that they may interfere
constructively giving rise to a unusually large intensity during some
coherence time $T$ of order $1/\Delta \omega $. In this case a transfer of
energy to the detection material will be most probable and an electron may
be ejected. It may be shown that the absorbed energy would be of order twice
the mean energy per mode, that is $E\sim 
\rlap{\protect\rule[1.1ex]{.325em}{.1ex}}h%
\omega ,$ see \cite{Foundations}.

\subsection{Linear momentum of the `photon'. Radiation needles}

In his 1916 work on emission of light by atoms, Einstein predicted that it
should be directional and random. The latter feature bothered Einstein by
the apparent violation of causality. The former was taken as a reinforcement
of the concept of photon, which acquired definite linear momentum in
addition to energy. Actually both features are straightforward consequences
of the ZPF. It is plausible that emission is estimulated by radiation either
sure or belonging to the ZPF. In the standard quantum language the former is
named stimulated and the latter spontaneous. The former would be in the same
direction than the incident beam, but the latter in a random direction due
to the stochasticity of the ZPF.

A more detailed, but semiquantitative, description of the emission is as
follows. Let us assume that a strong fluctuation of the ZPF with frequency $%
\omega $ arrives at an atom and it happens that $\omega $ is also one of the
possible frequencies for emission from the excited atom. Then the arriving
plane wave component of the ZPF may induce the emission of radiation with
the same frequency and phase than the incoming wave. The emitted radiation
should correspond to the addition of the amplitudes (not the intensities!)
of the incoming plane wave plus the emitted spherical wave. The frequencies
being equal there would be interference and it is not difficult to show that
it will be constructive in the forward direction and mainly destructive in
all other directions. The outgoing energy will be concentrated within the
region where the phase difference is small, with the boundary defined by the
following relation with the distance, $d$, and the half angle, $\theta $, as
seen from the atom. Then we have 
\begin{equation}
\frac{d}{\cos \theta }-d\sim \frac{\lambda }{2}\Rightarrow \theta \sim \sqrt{%
\frac{\lambda }{d}},  \label{needle}
\end{equation}
where $\lambda $ is the wavelength. If we take $d$ to be the coherence
length of the emitted light wavepacket (the alleged ``photon''), for typical
atomic emissions we have $d\sim 1$m, $\lambda \sim 1\mu ,$ so that $\theta
\sim 10^{-3}$. This fits with Einstein\'{}s proposal of ``needles of
radiation'' in the atomic emmison.

\subsection{Compton effect}

As is well known Compton\'{}s was the experiment that the
scientific community accepted as the final proof of the existence of
photons. The experiment is usually understood as a collision between one
photon of X-ray, with frequency $\omega _{1},$ and one electron, giving rise
to another photon with smaller frequency, $\omega _{2},$ at an angle $\theta 
$ with the incident radiation and a recoil electron. Indeed the
(relativistic) kinematics may be explained assuming that there are incident
and outgoing radiation needles having energies $
\rlap{\protect\rule[1.1ex]{.325em}{.1ex}}h%
\omega _{1}$ and $
\rlap{\protect\rule[1.1ex]{.325em}{.1ex}}h%
\omega _{2},$ respectively, and the electron is initially at rest. In
summary quantum electrodynamics (in the canonical formalism) gives a
quantitative account of the phenomenon, including the cross section of the
process, but it does not offer a clear intuitive picture. The WW formalism
for the field, in particular the random stochastic vacuum field, provides a
stochastic picture if we substitute radiation needles for photons \cite
{Foundations}. However the derivation of the cross section in WW seems
involved and will not been attempted here.

\subsection{Anticorrelation and recombination experiment}

A remarkable experiment showing the particle behaviour of light is the
anticorrelation after a beam splitter \cite{Grangier}. In the experiment%
\emph{\ }a light beam is sent to a balanced non-polarizing beam splitter
BS1. Two detectors, say A and B, placed in front of the two outgoing
channels may measure the single, P$_{A}$ and P$_{B}$, and coincidence, P$%
_{AB}$, detection probabilities within a small time window. We expect that
the probabilities fulfil 
\begin{equation}
r\equiv \frac{P_{AB}}{P_{A}P_{B}}\simeq \frac{\left\langle
I^{2}\right\rangle }{\left\langle I\right\rangle ^{2}},  \label{BSr}
\end{equation}
assuming that the probabilities are proportional to the intensities arriving
at both detectors, these supposed identical. The measured rates are given by
the products of detection probabilities times the number of windows in a
unit time interval.

If the radiation has a sure (nonfluctuating) intensity, like in a laser
beam, then we have 
\[
\left\langle I^{2}\right\rangle =\left\langle I\right\rangle ^{2}\Rightarrow
r=\frac{\left\langle I^{2}\right\rangle }{\left\langle I\right\rangle ^{2}}%
=1 
\]
meaning that the detections are uncorrelated. On the other hand for chaotic
(e.g. thermal) light we would have 
\[
\left\langle I^{2}\right\rangle =2\left\langle I\right\rangle
^{2}\Rightarrow r=\frac{\left\langle I^{2}\right\rangle }{\left\langle
I\right\rangle ^{2}}=2. 
\]
The change from $r=1$ to $r=2,$ a phenomenon known as ``photon bunching'',
has been interpreted as a quantum effect due to the Bose character of
photons. But a simple classical explanation is that it derives \emph{from
correlated Gaussian} fluctuations of the chaotic light.

The particle behaviour of light appears if the radiation incoming BS1 is
weak, and also the set up is appropriate in order to prepare the beam as a
series of `single photon states' in quantum language. In this case quantum
theory predicts, and the experiment confirms \cite{Grangier}, that the value
of $r$, eq.$\left( \ref{BSr}\right) ,$ is much smaller than unity. The
current quantum explanation is that \emph{photons are not divided}, but go
to one of the channels each. However in the WW formalism the particle
behaviour is caused by the ZPF \cite{MS}, \cite{Foundations}. In fact in
BS1, in addition to the signal entering one incoming channel, there is
another incoming channel where ZPF may enter, which interfere with the
signal beam. The interference should be destructive in one of the outgoing
beams if it is constructive in the other one, by conservation of energy. If
we assume that radiation is detected only when it is more intense than the
average ZPF beam, then there may be detection only in one of the detectors,
so explaining the anticorrelation.

Grangier et al.\cite{Grangier} showed also a wave behaviour of light in the
recombination experiment, where the detectors in front of BS1 are removed,
and via appropriate mirrors the two beams emerging from BS1 are send to two
incoming channels of another beam splitter BS2. The intensity emerging from
one of the outgoing channels of BS2 corresponds to the superposition of the
beams arriving to the incoming channels. The result of the experiment is
that detection is observed only in one the detectors placed in front of the
outgoing channels of BS2, but which detector clicks depends on the
difference between the two path lengths in the travel of light between BS1
and BS2. The standard quantum explanation is that light behaves as a wave
and the recombination reproduces the initial beam sent to BS1.

The recombination may be also easily interpreted within the WW formallism,
but I omit the details \cite{MS}, \cite{Foundations}.

Anyway the current interpretation is that the experiments show both particle
and wave behavior or light. The former in the anticorrelation and the later
in the recombination. Thus experiment is mentioned in popular books as a
proof of the impossibility of a realistic interpretation of quantum
mechanics.

\section{Entanglement and Bell inequalities}

\subsection{Entangled photon pairs from parametric down conversion}

Entanglement is a quantum property that may be easily defined mathematically
within the canonical (Hilbert space) formalism, but the definition does not
provide an intuitive picture of the phenomenon. It is currently seen as a
specific quantum form of correlation, which is claimed to be dramatically
different from the correlations that appear in classical physics. However
the WW formalism provides a picture for entangled ``photon pairs'', at least
those produced via ``spontaneous parametric down conversion'' (SPDC), which
is the most widely used method to get entangled photon pairs. The production
of ``photon pairs'' via SPDC is as follows \cite{Kaled}.

A laser beam with frequency $\omega _{0}$ is sent to an appropriate crystal
possesing nonlinear electric susceptibility. Then an a rainbow appears in
the opposite side of the crystal. Typically two (``conjugated'') beams $A$
and $B$ are selected amongst those in the rainbow, with frequencies $\omega
_{a}$ and $\omega _{b}$ fulfilling 
\[
\omega _{a}+\omega _{b}=\omega _{0}, 
\]
We may assume that a light beam with (complex) amplitude $a,$and frequency $%
\omega _{a},$ from the ZPF, enters the crystal on the same side than the
laser. Then the interaction of the laser, the beam $a$ and the electrons of
the crystal give rise to radiation of light with amplitude $a^{*}$ in a
different direction. In the same direction than $a^{*},$ a beam with
amplitude $b,$ also from the ZPF, enters the crystal and its interaction
with the laser and the electrons produces a radiation with amplitude $b^{*}$
and frequency $\omega _{b}.$ It is the case that this beam travels in the
same direction than $a.$ The result is that two rays emerge from the crystal
and they may be represented as functions of time $t$, by 
\begin{equation}
A\left( t\right) =a\left( t\right) +Db\left( t\right) ^{*},B\left( t\right)
=b\left( t\right) +Da\left( t\right) ^{*},  \label{entan}
\end{equation}
where D is a complex parameter, $\left| D\right| <<1.$ The direction of the
outgoing beam A is the same as that of the incoming beam $a$ and the
produced beam $b^{*}$, and similarly B, $b$ and $a^{*}$ are colinear. Of
course other radiation from the ZPF may enter the crystal giving rise to
other beams which are not collected in the experimental set up. The beams
actually selected in the experiment are the outgoing $A$ and $B$. For a
derivation within classical electrodynamics (but taking the ZPF as real) see
Ref.\cite{Kaled}.

The calculation within the Weyl-Wigner formalism of the correlation between $%
A\left( t\right) $ and $B\left( t\right) $ agrees with the predictions of
the canonical formalism for the correlation experiment. However the common
interpretation of the canonical calculation is that the beams $A$ and $B$
consist of pairs of photons, one photon in each beam, which are entangled.
Our analysis within WW provides an intuitive picture where the strong
correlation amongst the beams A and B comes from the fact that the beams
have fluctuating intensity and the positive fluctuations of $a^{*}$ coincide
in time with those of $a$. Similarly for $b$ and $b^{*}.$ As a consequence
there will be also correlated fluctuations amongst A$\left( t\right) $ and B$%
\left( t\right) $. Assuming that detection events happen at those times when
the beams arriving at the detectors have high intensity, then coincidence
detections may happen most frequently when fluctuations coincide. The
relevant result is the prediction that the coincidence detection rate $%
R_{AB} $ may close to the single detection rates. That is 
\begin{equation}
R_{A}\simeq R_{B}\simeq \ R_{AB},  \label{entancorr}
\end{equation}
to be compared with typical classical (without ZPF) where $%
R_{AB}<<R_{A}\simeq R_{B}$. This analysis within WW provides an intuitive
picture of entanglement \cite{Foundations}, \cite{Frontiers}, \cite{FOOP}.

In the canonical (Hilbert space) derivation the result involves the
commutation relations amongst creation and annihilation operators, but it
does not provide any intuitive picture. In the WW the ``photon
entanglement'' appears as a \emph{correlation between the fluctuations of
the field intensities of two beams}. The canonical (Hilbert space)
counterpart of eq.$\left( \ref{entan}\right) $ is 
\begin{equation}
\hat{A}=\hat{a}+D\hat{b}^{*},\hat{B}=\hat{b}+D\hat{b}^{*}.
\end{equation}

\subsection{Local realism and Bell inequalities}

As is well known Bell derived in 1964 some inequalities which should be
fulfilled by any local hidden variables (LHV) model. Later the concept of
LHV was generalized and the Bell inequalities are currently assumed to be
valid for any realistic local model. Thus the inequalities are used in order
to discriminate between quantum theory and local realistic theories. It is
assumed that any empirical violation of a Bell inequality would imply ``the
death of local realism'' .

Many experimental tests of the inequalities have been performed, those
involving ``entangled photon pairs'' being the most relevant. The current
opinion is that in fact local realism has been empirically refuted \cite
{Wise}. However I believe that the subject is not yet closed, but this
belief would require careful and long arguments in order to be convincing.
Thus the matter will not be studied further in this article. My arguments
may be seen in the references \cite{FOOP25} \cite{EPJP1}\emph{. }

\section{Conclusions}

I contend that physics should provide a coherent account of reality, in
addition to offer an algorithm for the prediction of empirical results.
However the mainstream of the physicists community believes that quantum
mechanics does not allow a realistic interpretation. In particular quantum
fields allegedly show a simultaneous behaviour as particles (localized) and
waves (extended), in spite of they being contradictory concepts. This fact
has given rise to a variety of interpretations of quantum mechanics, none of
which has reached a consensus. The Copenhagen interpretation is still the
most popular, but in fact it is rather the statement that no interpretation
is needed, just a good predictive power is required for physical theories. I
do not agree, and I have made efforts to get a ``realistic'' view of nature
which reaches consensus as it happens in classical physics.

My efforts have had only partial success, as is reviewed in this article.
The most relevant achievement is the proof that the Wigner (or Weyl-Wigner,
WW) representation predicts for Bose quantum fields the same (correct)
results than the most common canonical formalism resting on Hilbert spaces.
The WW formalism leads to a straightforward realistic interpretation of the
Bose fields as continuous, that is wavelike. The most important result is
the existence of \emph{vacuum fields} in the form of Gaussian random
radiation (sometimes named the zeropoint field, ZPF). The study of these
results is made in section 3.

The existence of the ZPF allows a qualitative interpretation of many
empirical results attributed to the ``corpuscular behaviour (or nature)'' of
the fields. This is the subject of section 6.

The wave behaviour of particles like electrons, neutrons, atoms or molecules
in interference experiments may be explained by non-local forces. That is we
may assume that the force may be mediated by some (hidden, not well known)
fields spread out over long distances. A simple model is presented in
section 5 where the motion of non-relativistic particles is formulated as a
probability distribution of paths, which depend non-locally on the Fourier
transform of the potential. This allows a simple interpretation of particle
interference, in section 5. in particular that, in the two-slit experiment,
one slit may influence the motion of a particle placed near another slit
separated from the former by a macroscopic distance.

Finally I offer an intuitive picture of entanglement, in particular
entangled photon pairs in section 7, where also I briefly comment on the
Bell inequalities.

The treatment of classical relativity is standard, except that I include a
philosophical proposal for the interpretation of general relativity in
section 2.5, which is independent on the rest of the article. Quantum
gravity is treated in section 4, rather superficially because I do not have
a clear opinion about it (and lack sufficient familiarity with the subject).
But I suggest that spacetime might be quantized as a consequence of the
matter quantization via Einstein equation, whence quantized spacetime would
mean assuming the existence of a probability distribution of metrics.

My article has an important incompleteness because I cannot give any hint
for a realistic interpretation of Fermi fields. Also I touch but slightly
the important problems of spacetime singularities (in black holes and the
very early universe) and the alleged loophole-free empirical violation of
Bell inequalities. Anyway I hope that the article provides an advance in our
understanding of physical reality.

\section{Statements and Declarations}

\textbf{Funding. }This research received no external funding.

\textbf{Data Availability Statement: }No new data were created of analyzed
in this study.

Data sharing is not applicable to this article.

\textbf{Conflict of Interest: }The author declares no conflict of interest.


\begin{thebibliography}{99}
\bibitem{Laloe}  F. Lalo\"{e}: Quantum mechanics is routinely used in
laboratories with great success, but no consensus on its interpretation has
emerged. In Ref.3.
\end{thebibliography}
\end{document}